\documentclass[fleqn,usenatbib, useAMS]{mnras}

\usepackage{newtxtext,newtxmath}

\usepackage[T1]{fontenc}
\usepackage{ae,aecompl}

\usepackage{graphicx}	
\usepackage{amsmath}	
\usepackage{amssymb}	
\usepackage{multicol}        
\usepackage{bm}		
\usepackage{pdflscape}	
\usepackage{subfigure}
\usepackage{hyperref}
\usepackage{subfigure}
\usepackage[dvipsnames]{xcolor}

\title[On the Mass of the UDG VCC 1287]{On the Stellar Kinematics and Mass of the Virgo Ultra-Diffuse Galaxy VCC 1287}

\author[J. S. Gannon et al.]{
Jonah S. Gannon$^{1}$\thanks{E-mail: jgannon@swin.edu.au}, Duncan A. Forbes$^{1}$, Aaron J. Romanowsky$^{2,3}$, Anna Ferr\'e-Mateu$^{4,1}$,
\newauthor{Warrick J. Couch$^{1}$ and Jean P. Brodie$^{1, 2}$}
\\
$^{1}$Centre for Astrophysics and Supercomputing, Swinburne University, John Street, Hawthorn VIC 3122, Australia
\\
$^{2}$ University of California Observatories, 1156 High Street, Santa Cruz, CA 95064, USA
\\
$^{3}$ Department of Physics and Astronomy, San Jos\'e State University, 1 Washington Square, San Jose, CA 95192, USA
\\
$^{4}$ Institut de Ci\'encies del Cosmos (ICCUB), Universitat de Barcelona (IEEC-UB), Barcelona 08028, Spain
}

\date{Accepted XXX. Received YYY; in original form ZZZ}

\pubyear{2019}

\begin{document}
\label{firstpage}
\pagerange{\pageref{firstpage}--\pageref{lastpage}}
\maketitle

\begin{abstract}
Here, we present a kinematical analysis of the Virgo cluster ultra-diffuse galaxy (UDG) VCC 1287 based on data taken with the Keck Cosmic Web Imager (KCWI). We confirm VCC 1287's association both with the Virgo cluster and its globular cluster (GC) system, measuring a recessional velocity of $1116 \pm 2\ \mathrm{km\ s^{-1}}$. We measure a stellar velocity dispersion ($19 \pm 6\ \mathrm{km\ s^{-1}}$) and infer both a dynamical mass ($1.11^{+0.81}_{-0.81} \times 10^{9} \ \mathrm{M_{\odot}}$) and mass to light ratio ($13^{+11}_{-11}$) within the half light radius (4.4 kpc). This places VCC 1287 slightly above the well established relation for normal galaxies, with a higher mass to light ratio for its dynamical mass than normal galaxies. We use our dynamical mass, and an estimate of GC system richness, to place VCC 1287 on the GC number -- dynamical mass relation, finding good agreement with a sample of normal galaxies. Based on a total halo mass derived from GC counts, we then infer that VCC 1287 likely resides in a cored or low concentration dark matter halo. Based on the comparison of our measurements to predictions from simulations, we find that strong stellar feedback and/or tidal effects are plausibly the dominant mechanisms in the formation of VCC 1287. Finally, we compare our measurement of the dynamical mass with those for other UDGs. These dynamical mass estimates suggest relatively massive halos and a failed galaxy origin for at least some UDGs. 
\end{abstract}




\section{Introduction}

The term ultra-diffuse galaxy (UDG) was first coined by \citet{VanDokkum2015} in reference to 47 faint, diffuse galaxies discovered in the Coma cluster with low central surface brightness ($\mu(g,0) \geq 24\ \mathrm{mag\ arcsec^{-2}}$) and Milky Way-like sizes ($R_{e}>1.5\mathrm{kpc}$). Since this discovery, UDGs have been found not only residing in other clusters (e.g. \citealp{Mihos2015, Venhola2017, Janssens2017}), but also in less dense groups (e.g. \citealp{Merritt2016, Martinez-Delgado2016, Roman2017b, Muller2018, Forbes2019}) and field environments (e.g. \citealp{Bellazzini2017, Roman2017}).

Theorists have already proposed several formation mechanisms able to account for many of the observed properties of UDGs. Primarily, they seek to explain UDGs as galaxies that formed either through internal processes, such as high halo spin \citep{Amorisco2016} and stellar feedback \citep{DiCintio2017}, or external influences, like tidal heating/stripping \citep{Yozin2015, Carleton2018} leading to the large, diffuse objects we see today. Combinations of these mechanisms have also been considered (e.g. \citealp{Rong2017, Jiang2019}).

Since UDGs have dwarf galaxy like stellar masses and Milky Way like sizes, their formation scenarios are usually categorised in two broad groups (e.g. \citealp{Forbes2020}). One forms UDGs from normal dwarf galaxies that are `puffed up' by some mechanism. The other suggests UDGs form from more massive Milky Way like galaxy halos that have failed to produce a large stellar population causing us to observe a `failed galaxy'. No formation theory has yet explained all of their properties satisfactorily. Several UDGs have been discovered with properties so extreme as to not be adequately explained by any current UDG formation scenario: those that have unusually large halo masses (e.g. Dragonfly 44, DGSAT I; \mbox{\citealp{Lim2018}}; \mbox{\citealp{vanDokkum2019b}}; \mbox{\citealp{Martin-Navarro2019}}), having an overabundance of GCs \citep{vanDokkum2017}, having exotic chemical abundances \citep{Martin-Navarro2019}, those that are quiescent in field environments \citep{Martinez-Delgado2016, Papastergis2017} and a pair that may lack dark matter entirely (NGC1052--DF2, NGC1052--DF4 \citealp{vanDokkum2018, vanDokkum2019, Danieli2019}), although controversy exists on this last point \citep{Trujillo2019}. 

The `puffy dwarf'/`failed galaxy' dichotomy leads to a natural line of investigation where, in measuring the mass of a UDG's halo, we are able to infer information about its formation history. There are multiple methods available to determine these masses such as the established GC mass -- halo mass relation \citep{Spitler2009, Beasley2016, Harris2017}, weak lensing \citep{Sifon2018}, HI rotation curves (\citealp{Leisman2017}; \citealp{Spekkens2018}; \citealp{ManceraPina2019};\citealp{Sengupta2019}), X-ray stacking \citep{Kovacs2019} or measuring either GC or stellar velocity dispersions to infer a dynamical mass for the galaxy within a given radius \citep{Wolf2010, Alabi2016}. Necessary measurements are not always available; some UDGs are known to lack HI (e.g. \citealp{Sardone2019}), weak lensing calculations can be effected by tidal stripping as may be present in some UDGs \citep{Sifon2018}, and X-ray stacking cannot provide masses for lone UDGs. Additionally, although the GC mass -- halo mass relation is thought to be able to provide accurate halo masses, we are limited in our ability to detect GCs around UDGs in all but the local Universe. The relation is also generally untested at lower masses \citep{Forbes2018, Burkert2019}, limiting confidence in its reliability for UDGs. We must then rely on mass tracers and measurements of velocity dispersion to infer dynamical masses for individual UDGs\footnote{Extrapolation of a dynamical mass into a total halo mass requires the non-trivial assumption of a halo profile.}. Note that detecting mass tracers such as GCs can suffer from similar issues of detection but have already been used to infer halo masses for several UDGs \citep{Beasley2016, Emsellem2018, Toloba2018, vanDokkum2018, vanDokkum2019}. 

In order to test the accuracy of these measurements ideally one would like to get multiple mass estimations for single objects and check their consistency. Due to their inherently faint nature, measuring a stellar velocity dispersion can be a particularly burdensome task requiring extremely long integration times on the largest telescopes and as such, has only been achieved for a handful of UDGs \citep{vanDokkum2016, Martin-Navarro2019, Emsellem2018, Danieli2019, vanDokkum2019b}. Only one of these galaxies, NGC1052--DF2, has associated GC kinematic mass measurements which were tested for consistency with those from its stellar velocity dispersion \citep{Danieli2019}. Evidence exists that this UDG does not have the elevated dark matter content known to exist in other UDGs \citep{vanDokkum2018a, Danieli2019, Trujillo2019}. We would therefore also need to examine an example of a high dark matter content UDGs in a similar manner.

It should be noted that large, diffuse, low surface brightness objects have been observed for decades and a subset have only recently been dubbed by the term UDG. For example UDG candidate VCC 1287 was tabulated as a dwarf "... of very large size and low surface brightness" in \citet{Binggeli1985}. In this work we analyse the stellar kinematics of VCC 1287 using observations taken with the Keck Cosmic Web Imager (KCWI) \citep{Morrissey2018}. VCC 1287 has been an important object in the field of UDGs as it was one of the first objects to have a mass inferred \citep{Beasley2016}. Through examination of GC dynamics and numbers \citet{Beasley2016} inferred a large virial mass and mass to light ratio ($M/L$) for the object implying a dark matter dominated system. Crucially the estimates used: tracer mass estimation (TME), GC number counting, and GC velocity dispersion, were all found to be consistent with one another \citep{Beasley2016}. Already having these properties measured we seek a recessional velocity for the stellar light to ensure it is indeed associated with the GC system discussed in \citet{Beasley2016} along with a measurement of velocity dispersion for the stars to perform complementary stellar kinematic analysis. If the stars are associated with the GCs we will then be able to compare the kinematics of the stellar body with those of the GCs, ensuring consistency between these mass estimation methods. We simultaneously seek to further constrain the halo mass of VCC 1287 using the new stellar kinematic measurement.

Due to the lack of optimisation in the reduction pipeline of the publicly available CFHT data to detect low surface brightness objects, an over-subtraction issue led to imperfect photometry being used in \citet{Beasley2016} (see appendix C of \citet{Pandya2018}). We therefore choose to use the updated VCC 1287 photometry in \citet{Pandya2018} throughout this work when deriving properties for VCC 1287. In Section \S 2 we summarise the acquisition, processing and initial reduction of our data. We discuss problems involving offset sky subtraction for diffuse objects in Section \S 3, presenting a modern method for the removal of sky flux. Section \S 4 uses our reduced spectrum to analyse the stellar kinematics of VCC 1287, discussing these results in a wider context. We present concluding remarks in Section \S 5. All magnitudes are quoted in the AB system unless otherwise stated and a Hubble constant $H_{0}$ of $70\ \mathrm{km\ s^{-1}\ Mpc^{-1}}$ is assumed. When using literature values for half light radius where errors were not quoted we assume an error of 10\%. For error propagation we exclude terms higher than second order in the Taylor expansion.

\begin{figure}
    \centering
    \includegraphics[width = 0.45 \textwidth]{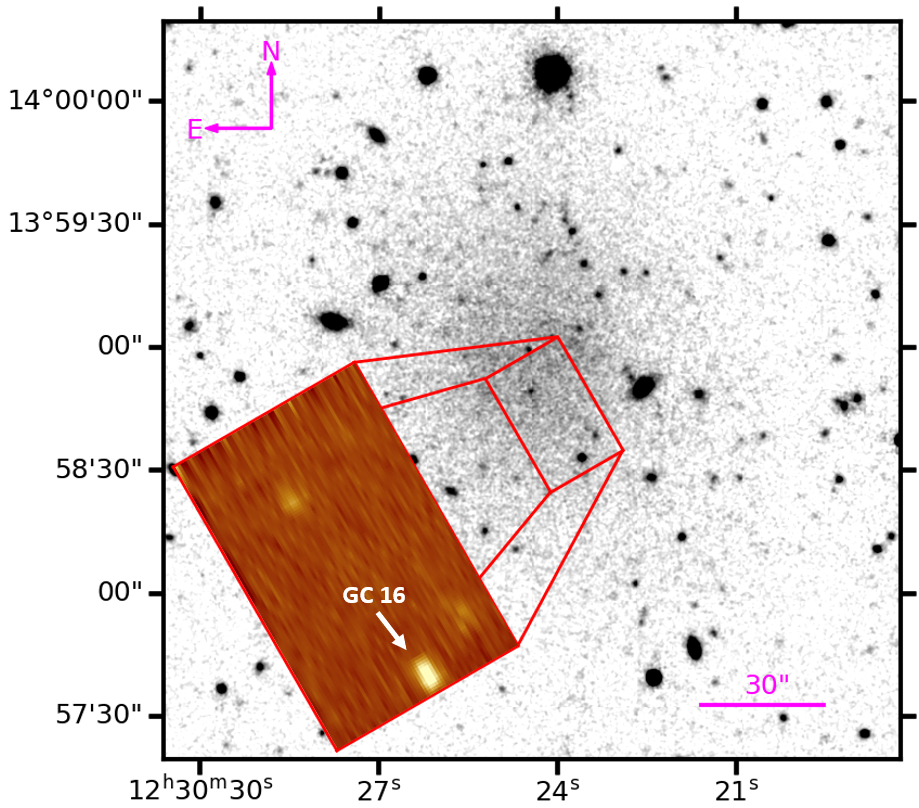}
    \caption{A 3' by 3' g-band cutout of VCC 1287 taken from the DECaLS DR8 sky-viewer\protect \footnotemark \ overlaid with the positioning of the 33" by 20.4" large slicer of KCWI. A pop-up of a single reduced, wavelength collapsed 1200s KCWI exposure is included. We have also labelled the compact object, GC16 ($m_{g}\ =\ 22.3$ mag), from \citet{Beasley2016}.}
    \label{fig:1287}
\end{figure}

\protect \footnotetext{\url{http://viewer.legacysurvey.org/?ra=187.6029&dec=13.9814&zoom=14&layer=dr8}}

\section{Data}

\subsection{Acquisition} \label{sec:acquisiton}
Observations were made during the first half of the nights 2019, March 29th and April 2nd. Due to the low surface brightness nature of VCC 1287 we used the large slicer with a 33" by 20.4" field of view and the BH3 grating. We set a central wavelength of 5080 \AA\ based on the assumption that the galaxy was associated with the GC system described in \citet{Beasley2016}. Under this assumption the resulting spectral coverage (4823\AA\ - 5315\AA) is such that it will create a wavelength buffer to our expected $\mathrm{H\beta}$ absorption line while also allowing the measurement of the $\mathrm{Mgb}$ triplet. In this configuration we characterise the instrumental resolution by fitting Gaussians to spectral features in: the arc-bar calibration files, sky emission lines in our offset sky exposures and absorption lines in a high S/N exposure of M3 observed in the same configuration on 2019, April 2nd. All three provide an instrumental resolution within 1\% of each other. From these measurements we adopt an instrumental resolution of $\sigma_{instrumental} = 25\  \mathrm{km\ s^{-1}}$. We note that formally the line spread function of KCWI may be non-Gaussian \citep{vanDokkum2019b} and will have a wavelength dependence. Our approximation of this profile as Gaussian is not expected to effect final results \citep{Emsellem2018}. We further consider our assumption of a fixed instrumental resolution without wavelength dependence in Section \ref{sec:RV+SK}.

Since our target fills the field of view of KCWI (see Fig.\ref{fig:1287}) we opted to intersperse dedicated sky exposures between our science frames to allow proper sampling of the sky emission. On 2019, March 29th we took object (O) and sky (S) exposures in an $S-S-S-S- 4 \times (O-O-S)$ pattern. The initial four sky exposures were taken to be 300s each in order to allow an investigation of the changing night sky emission over the period when it is expected to vary most, immediately following twilight. All other exposures were 1200s. On 2019 April 2nd four additional exposures were taken in an $S-O-O-S$ pattern with both sky exposures being of 600s length and both science exposures being of 1200s length. We also took a 300s exposure on both the Milky Way GC M3 and a nearby offset sky region in order to perform the instrumental resolution investigation. Total exposure times were 3 hrs 20 mins on VCC 1287, 2 hrs on nearby sky, and 5 mins each on M3 and its offset sky.

\subsection{Processing and Spectrum Extraction}
The data were reduced using the standard KCWI data reduction pipeline \textit{KDERP} \citep{Morrissey2018} with minor alterations described below. For all the following science we use the non-sky subtracted, standard star calibrated, `ocubes' files created by the reduction pipeline unless otherwise specified. In order to correct for a low level gradient observed in the reduced data cubes we perform an additional correction as described in Appendix \ref{app:flat} (see also section 3.2 of \citealp{vanDokkum2019b}). We then crop the data cubes, extracting spectra and performing sky subtraction as per Section \S \ref{sec:skysub}. Data cropping is necessary to remove the spatially padded borders of the data cube introduced by the pipeline, along with outer spaxels with systematically offset flux values. Additionally, we read the `WAVGOOD0' and `WAVGOOD1' header terms of the data cubes in order to crop the cubes to the wavelength range which all slices have coverage.

The UDG spectra are extracted from the data cubes by selecting all spaxels except a three by ten spaxel box surrounding the GC, GC16 from \citet{Beasley2016}, in our data cube. We collapse these spaxels in light-bucket mode to create a single spectrum for each observation of the UDG. For our sky frames we perform the cube cropping and flat fielding correction using the same python scripts. We extract the sky spectra by selecting the spaxels with flux within two median--absolute--deviations of the median in the wavelength collapsed data cube and mean stacking their spectra to produce a single spectrum for each data cube. These cuts ensure any faint objects accidentally included in these observations are not included in our extracted spectra. 

\section{Sky Subtraction} \label{sec:skysub}
\subsection{Building a Model} \label{sec:bam}

\begin{figure*}
    \centering
    \includegraphics[width = 0.75 \textwidth]{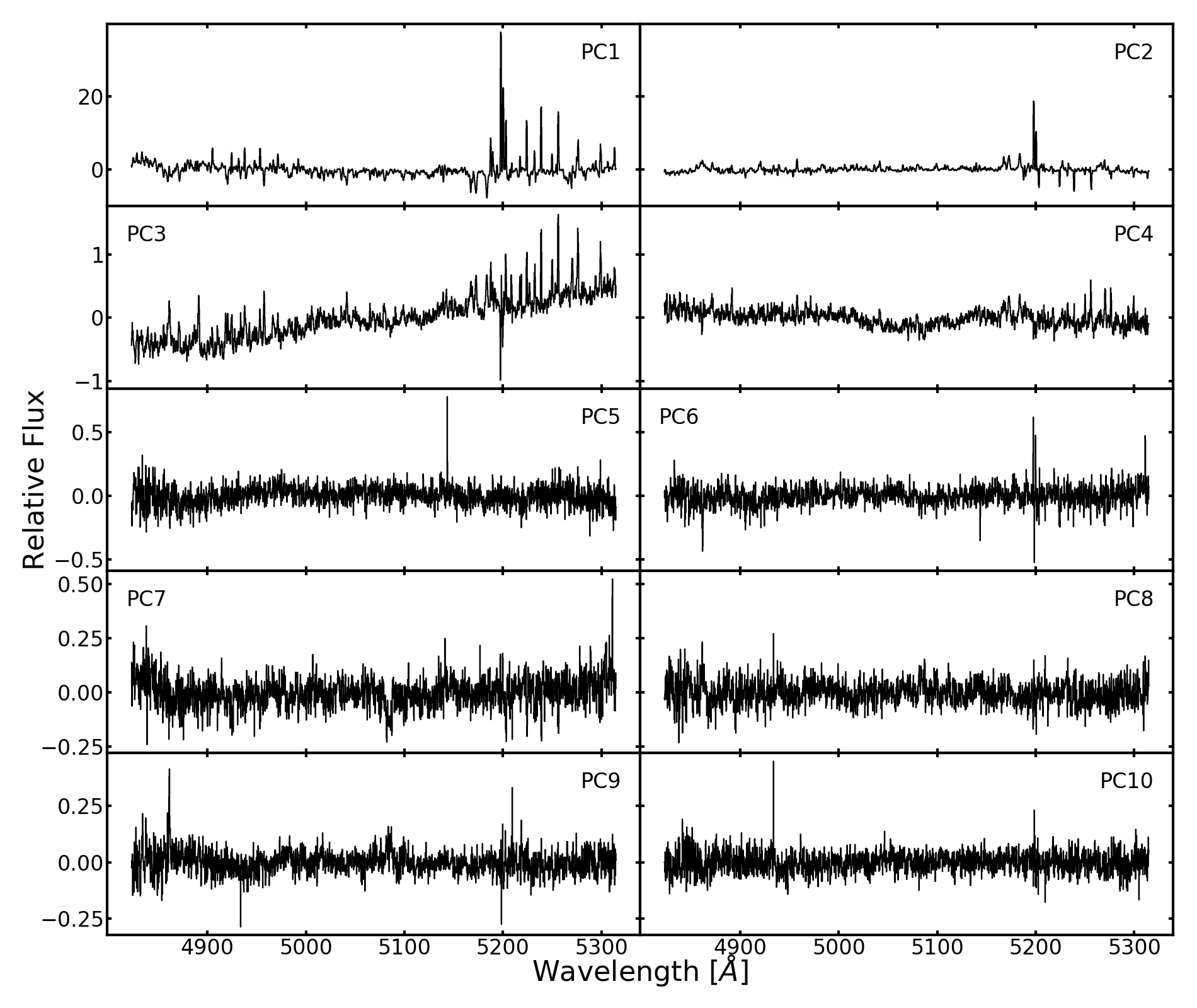}
    \caption{The 10 principal components we extracted from our offset sky frames. Of particular interest is the change in flux levels moving to higher principal components. Most of the spectral feature information is captured in the first four components, yet higher order components are still useful to ensure the continuum can be accurately recovered.}
    \label{fig:pcs}
\end{figure*}

The particular challenge with the data collected for this study is the accurate subtraction of the sky from the science exposures using non-contemporaneous offset sky exposures. Here, we require a manner to relate what we have observed in our offset sky exposures with the sky in our science frames. We initially adopted the method of principal component analysis (PCA) through use of Zurich Atmospheric Purge (ZAP - \citealp{Soto2016}) to decompose the temporally nearest sky frame. We found that, due to temporal variations in the sky, the contaminant emission in this frame is not representative of that in the science exposures, despite being observed directly after. The principal components (PCs) of the offset frame therefore cannot be used to reconstruct the sky in the science observation. Recent work by \citet{vanDokkum2019b} and \citet{Danieli2019} proved PCA to be a highly effective method for characterising the contaminant emission as it creates a basis from which a simple linear combination of its component eigen-spectra can be used to model the contamination. We thus write our own PCA sky subtraction program that examines not just the offset sky exposures temporally close to the relevant science frame but all sky exposures taken across both observing nights\footnote{The version used in this paper is available at: \\ \url{https://github.com/gannonjs/Published_Code/blob/master/VCC1287_PCA_sky_subtraction.py}}. In using the full ensemble of sky exposures to build our PCs we expect these components to now describe any temporal variations that exist in our data. We use these components to build a model of our observations expecting that the flux observed while targeting VCC 1287 will be some combination of contaminant emission from a myriad of other sources, all contained within the PCs, along with the flux of interest (VCC 1287).

Under the above assumption, that the contaminant emission ($\mathcal{C}$) is comprised of the same principal components ($\mathcal{PC}$), only with differing magnitudes ($\alpha$), in our sky and science exposures, we can build a model of the contaminant emission ($\mathcal{C}$) in our observations. This model is a simple linear combination of the components:

\begin{equation} \label{eqn:cont}
    \mathcal{C} = \sum^{n}_{i=1}\  \alpha_{i}\  \mathcal{PC}_{i}
\end{equation}

Having extracted a spectrum for each of our ten offset sky exposures across both observing nights, we decompose them into their PCs using the python package \textit{scikit--learn} \citep{scikit-learn}. This creates ten eigen-spectra that formed the basis for the temporal changes in the contaminant emission across our sky observations. 

To extend equation \ref{eqn:cont} to a model of our science observations, we add a component to account for emission coming from VCC 1287 along with a pedestal to balance differing continuum levels in our components. Not knowing the VCC 1287 redshift \textit{a priori} a further redshifting parameter for the galaxy template is also required. Our final model of observed emission is then $M(m_{1},\ m_{2},\ ...\ ,\ m_{13})$, where parameters $m_{1}$ through $m_{10}$ correspond to our contaminant emission linear components ($\alpha_{i}$). The parameters $m_{11}$, $m_{12}$ and $m_{13}$ are used to vary the strength of the galaxy template ($\mathcal{T}(z)$), provide a pedestal to account for simple continuum differences and adjust the redshift of the template, respectively:

\begin{equation}
    M(m_{1},\ ...\ ,\ m_{13}) = \sum^{10}_{i=1}\  m_{i}\  \mathcal{PC}_{i}\ + m_{11}\mathcal{T}(z = m_{13}) + m_{12}
\end{equation}

We display the ten PCs extracted for use in this model in Figure \ref{fig:pcs}. In practice using all ten components is unnecessary as the majority of the information is captured in the first four components and results do not vary significantly when using fewer components. We were limited to ten PCs by the number of offset sky frames we have. The first PC is noticeably similar to the average of all sky frames. Of particular interest are the visible solar absorption features in the spectrum despite the observations being taken during dark moon conditions. The second PC captures the variation in the emission lines around 5200 \AA\ along with the variation in the solar absorption. The third and fourth PCs capture variations in the continuum along with any variations in emission lines not captured by PC2. Beyond these components the information contained in the PCs mostly involves low level continuum and noise variations. As we process the data differently before running PCA (e.g. standard star calibration, instrumental configuration, etc. ) our PCs differ to those plotted in figure 3 of \citet{vanDokkum2019b} although they both capture much of the same variations. Our PC1 resembles the \citet{vanDokkum2019b} sky frame average. Our PC2 resembles their PC4 capturing the variation in the stronger OH lines, the solar absorption features and the [N I] emission around 5200 \AA\ and our PC3 similarly captures the OH emission variation in our spectra. 
 
In addition to these PCs we also require a template for our galaxy emission. Based on the optical and near-IR spectral energy distribution fitting for VCC 1287 in \citet{Pandya2018}, we expect our stellar population to be intermediate to old ($\ge 7.5$ Gyr) and metal poor ([Z/$\mathrm{Z_{\odot}}$] < $-1.55$) \citep{Pandya2018}. Spectroscopic studies of other UDGs yield similar results for their stellar population \citep{Gu2018, Ferre-Mateu2018, Ruiz-Lara2018}. For this reason, we selected twelve high resolution templates from the \citet{Coelho2014} library of high resolution synthesised stellar spectra and smoothed them to the resolution of KCWI. Nine templates were chosen for their resemblance to K-type giants (-0.3 < [Fe/H] < -1.3\footnote{[Fe/H] = -1.3 is the lowest metallicity in the \citet{Coelho2014} library.}) as it has long been known these stars are representative of the type of old, metal poor population we expect to observe (e.g. \citealp{Morgan1957}). The remaining three templates were chosen to resemble an A, F and G spectral type star, respectively. We performed fits with all to test the effect of template mismatch on our best fitting model. As expected the K-type giant templates consistently provided the best fits to our data. Encouragingly, when fitting all templates we found a similar redshift parameter for VCC 1287 in our model. In the case of using the A type stellar template for galaxy emission the best fit for the parameter of its normalisation is sufficiently small to be essentially unused in the sky subtraction. In such a case, where the galaxy template is not used, we were still able to recover many of the same spectral features. For the remaining reduction we select one of these K-giant templates ([Fe/H] = -1) as our template for galaxy emission. 
 
 \subsection{Fitting the Model} \label{sec:fam}

 \begin{figure*}
    \centering
    \includegraphics[width = 0.75 \textwidth]{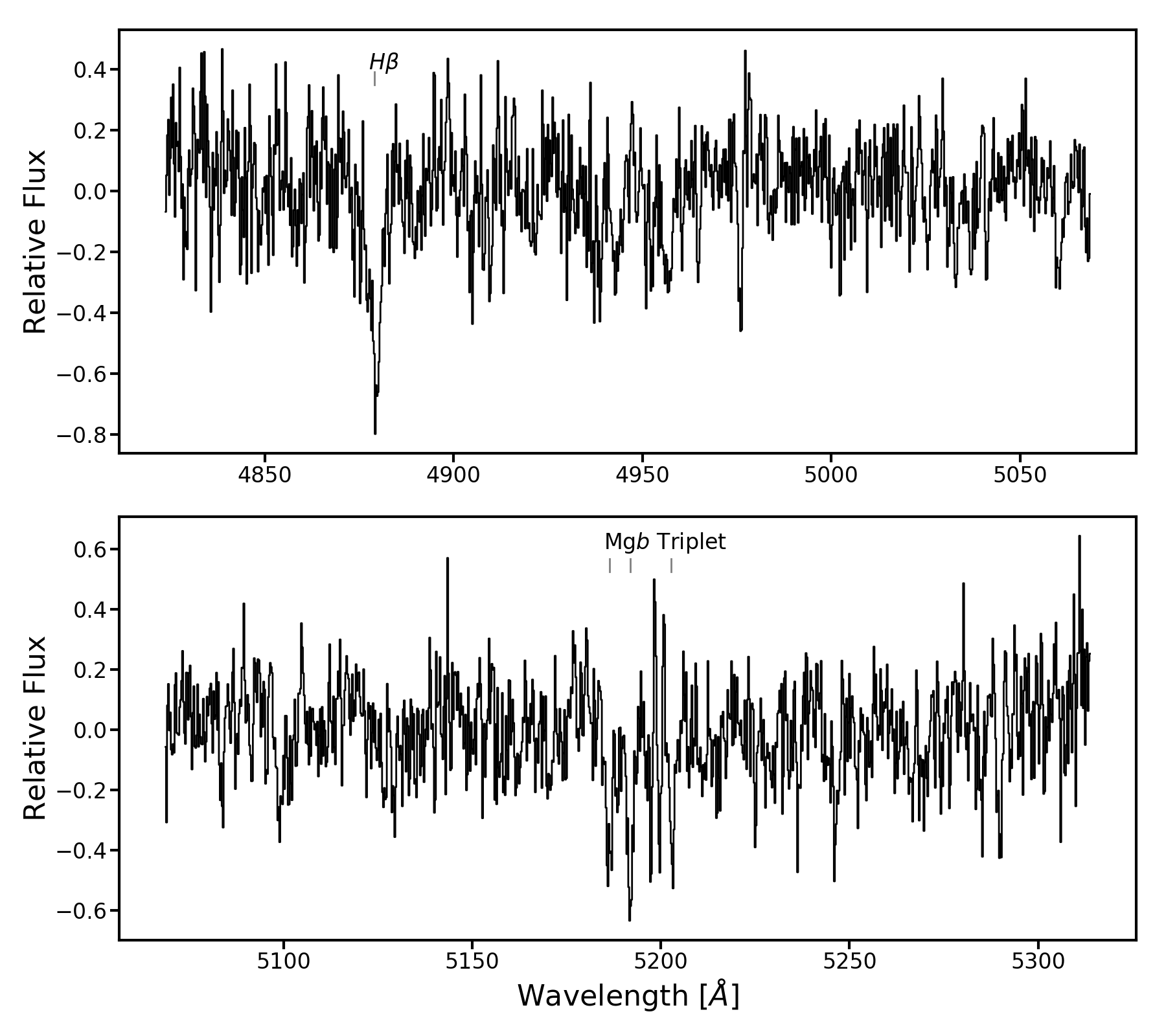}
    \caption{The final stacked spectrum we use for deriving the internal kinematics of VCC 1287 after sky subtraction. Barycentric corrections were applied before stacking. For ease of viewing, the spectrum has been split into two halves. $\mathrm{H\beta}$ and the $\mathrm{Mgb}$ triplet have been labelled.}
    \label{fig:SSspectra}
\end{figure*}

\begin{figure*}
    \centering
    \includegraphics[width = 0.9 \textwidth]{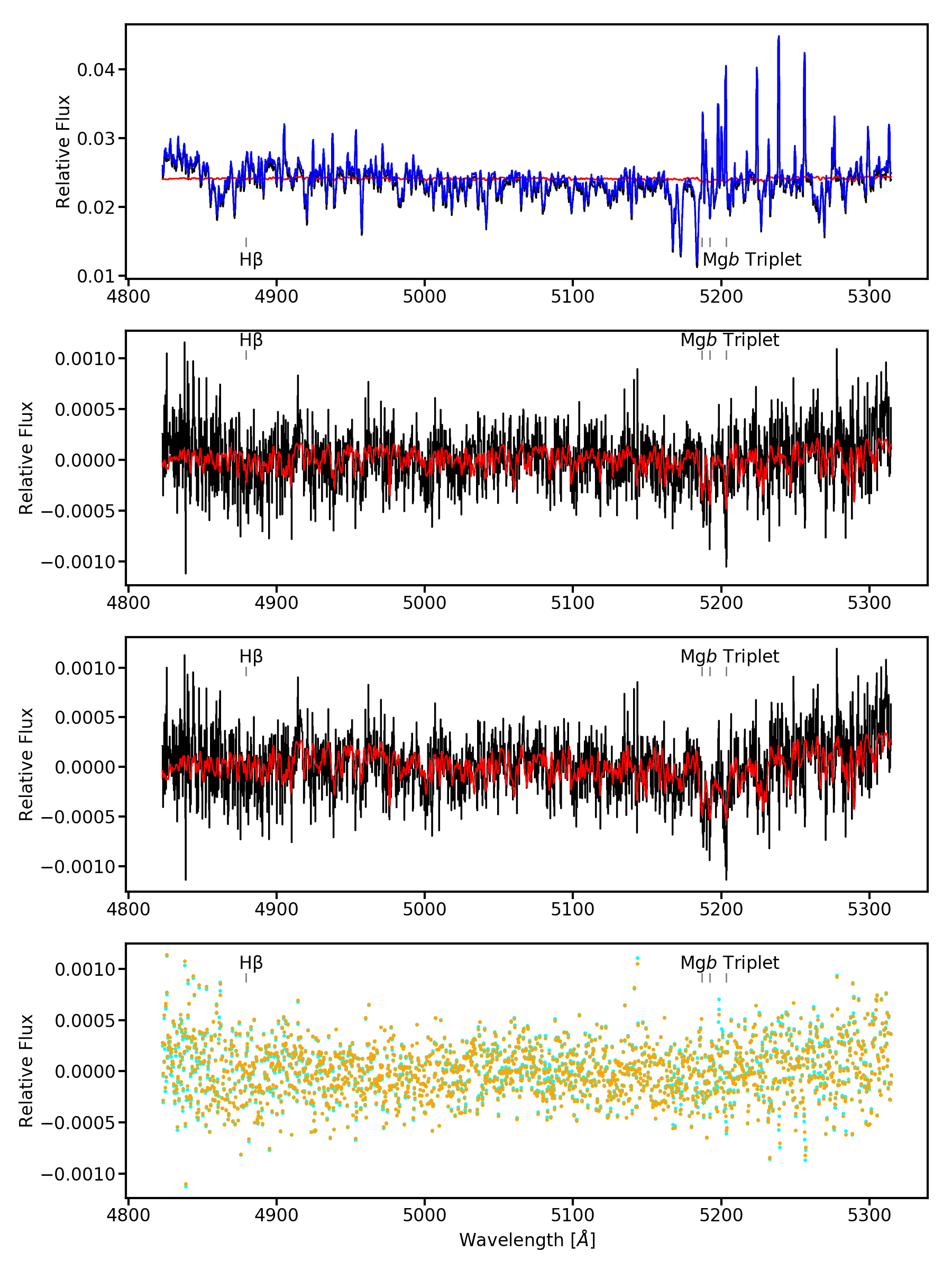}
    \caption{The creation and subsequent sky subtraction of our mock data. \textit{Upper Panel}: Our mock spectrum. Galaxy template emission (red with added pedestal) is added to a sky spectrum (black) to create our mock data (blue). \textit{Second Panel from Top}: A comparison between our sky subtracted mock data set (black) and the template inserted into the data (red). \textit{Second Panel from Bottom}: The same as the second panel from the top but inserting a differing template into the data to what is used in the model of the observation. \textit{Bottom Panel}: the residuals for the extraction using the same template (orange) and differing templates (cyan). In all panels the positioning of $\mathrm{H\beta}$ and the $\mathrm{Mgb}$ triplet in the galaxy template has been labelled. Our sky subtraction routine is able to effectively recover flux in our mock data.}
    \label{fig:mock}
\end{figure*}

Having built a model of what we expect to have observed, we then fit it to our data using a maximum likelihood estimator. In performing this maximisation we find a small dependence of the output parameters on our best guess input parameters. Due to this we add an additional step where we take this output parameter set and use it to initialise Markov Chain Monte Carlo (MCMC) analysis of our posterior distribution using \textit{EMCEE} \citep{EMCEE}. We allow for broad uniform priors. Investigations show burn in occurs in our sample after $\sim$ 700 iterations although we only take samples after 800 iterations to be certain we are sampling the posterior distribution. To ensure we explore the entire parameter space adequately we double the number of walkers until we have confidence we are fully exploring the posterior distribution. We use 800 walkers. Posterior distributions are Gaussian and independent for all parameters. We take the mean value of these distributions as the optimal parameters in constructing the final model of our data. If we use the median value of these posterior distributions in constructing the model we find no noticeable difference. Subtracting the sky component of the optimal model from our data we arrive at our sky subtracted science frames. We apply the relevant barycentric corrections using \citet{Tollerud2013} and median combine the nine frames to create our final science spectra. We show the resulting median combined spectra in Figure \ref{fig:SSspectra}. We estimate a signal to noise of 17 per \AA\ at H$\mathrm{\beta}$ for this spectrum.

In order to check that our sky subtraction process is able to correctly recover the galaxy flux, we build two mock data sets. For both we take one of our sky exposures and insert flux from one of two K-type giant synthetic stellar templates from \citet{Coelho2014} as per Figure \ref{fig:mock}. One of these templates is the same as that which we use to reduce our data, the other is simply a similar K-giant we use to demonstrate our subtraction is robust in the case of slight template mismatch. We show the results of running our sky subtraction on both of these data sets (without the sky frame used to create them) in Figure \ref{fig:mock}. When the template inserted into our mock data matches the template we use in our model for the observation, we are able to recover our inserted flux with additional noise. In the case of a slight mismatch between the template in our model and the `galaxy' flux in our mock data we are still able to recover the flux we have inserted with a similar level of added noise. We conclude our sky subtraction is therefore able to effectively recover our artificially inserted flux, giving confidence in our extracted science spectra, even in the likely scenario that our recovery template does not perfectly match the actual flux from VCC 1287.

\section{Results and Discussion}

\subsection{Recessional Velocities and Stellar Kinematics} \label{sec:RV+SK}

\begin{figure*}
    \centering
    \includegraphics[width = 0.85 \textwidth]{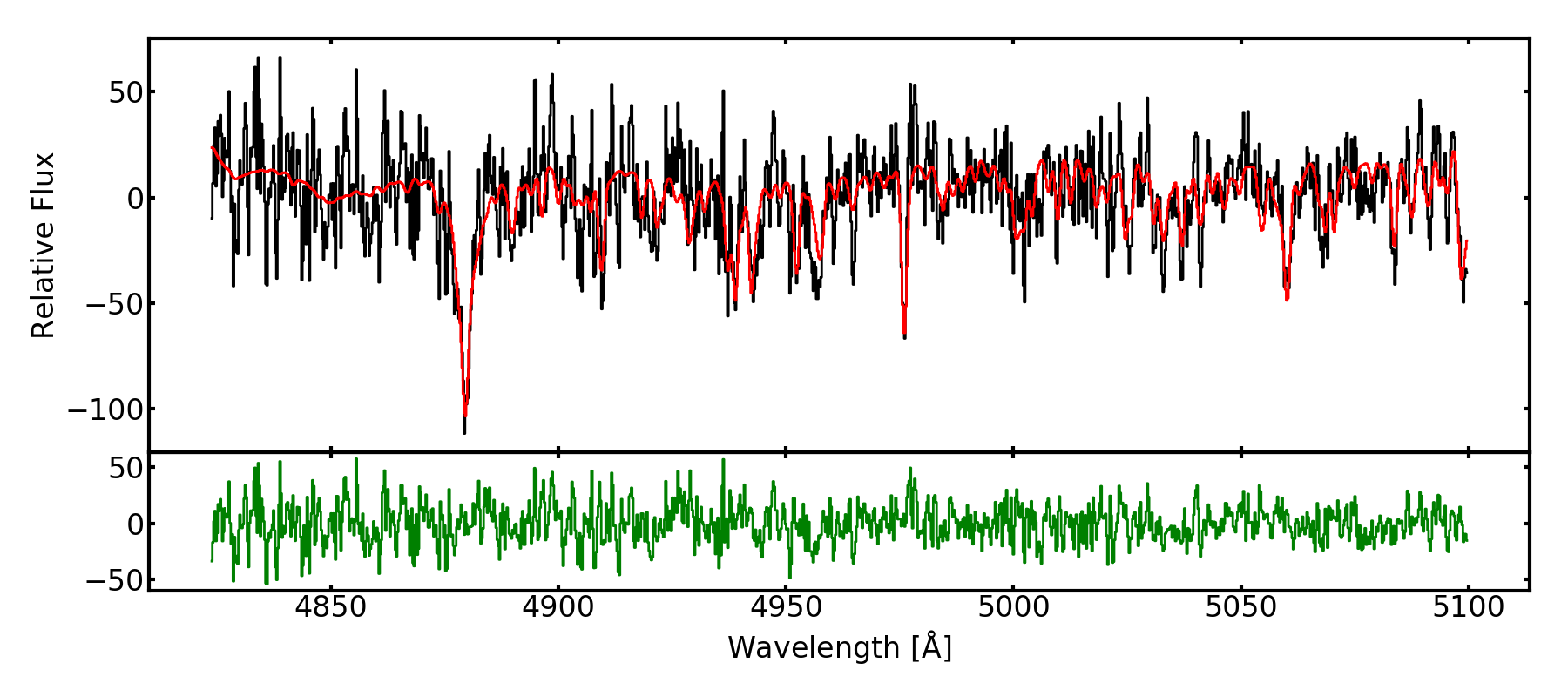}
    \caption{A representative VCC 1287 \texttt{pPXF} spectrum fit. \textit{Upper Panel}: Our science spectrum (black) with \texttt{pPXF} best fitting template (red; from the \citealp{Coelho2014} library) used to extract a recessional velocity and kinematics. \textit{Lower Panel}: The residuals of the fit (green). Based on this fit we measure a recessional velocity of $1116 \pm 2$ $\mathrm{km\ s^{-1}}$ and a velocity dispersion of $16 \pm 4$ $\mathrm{km\ s^{-1}}$. The spectrum has been cropped to the fitting region.}
    \label{fig:pPXF_All}
\end{figure*}

\begin{figure}
    \centering
    \includegraphics[width = 0.47 \textwidth]{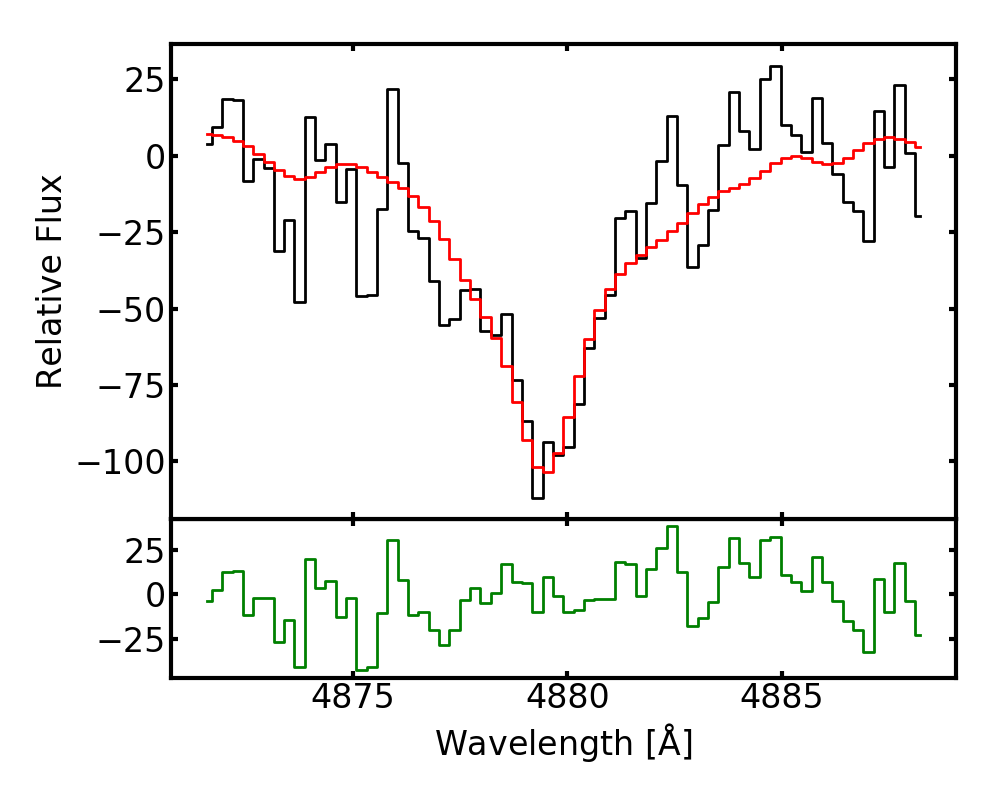}
    \caption{VCC 1287 \texttt{pPXF} $\mathrm{H\beta}$ fit. \textit{Upper Panel}: A zoomed version of Figure \ref{fig:pPXF_All} in the region of our science spectrum around $\mathrm{H\beta}$ (black) we fit with \texttt{pPXF} to extract a recessional velocity and velocity dispersion along with best fitting template (red). \textit{Lower Panel}: The fit residuals (green).}
    \label{fig:pPXF_Hb}
\end{figure}

In processing our data we extracted two estimates of the recessional velocity of VCC 1287. The first we obtain from the parameter used for the red-shifting of our galaxy emission template during the sky subtraction process. The second we obtain from running \texttt{pPXF} \citep{Cappellari2017} on our final spectra.

In the first case we extract a redshift for the template by taking the mean of the relevant MCMC posterior distribution parameter and transform it into a recessional velocity. We then apply a barycentric correction calculated using the publicly available code from \citet{Tollerud2013}. The mean of our barycentric corrected recessional velocities for each of the nine science frames extracted from the MCMC posterior distributions is $1114$ $\mathrm{km\ s^{-1}}$ with a standard deviation of $4$ $\mathrm{km\ s^{-1}}$.

Using \texttt{pPXF} we fit only the region of our spectra blue-ward of 5100\AA\ after logarithmic rebinning in order to avoid the region worst affected by sky residuals. This procedure excludes all of the $\mathrm{Mgb}$ triplet. In the case where sky residuals are accidentally included in our fit we would be biased towards lower velocity dispersions. In the case where VCC 1287 has experienced an atypical star formation history (i.e. high [Mg/Fe]), as may be expected for UDGs \citep{Martin2019}, template mismatch in the $\mathrm{Mgb}$ triplet is known to bias fitting of velocity dispersion \citep{Barth2002}. As a test, if we include this region in our fitting and instead fit the entire spectrum we find our velocity dispersion is biased towards lower values.

We fit our spectra using two different methods. First, using the \citet{Coelho2014} stellar library and a wide range of input parameters for \texttt{pPXF}, we discard the <2 \% of fits we deem to to have ineffectively modelled the data (see Appendix \ref{app:fitting}). Taking the median of the subsequent distributions for each parameter we recover a recessional velocity of 1116 $\pm$ 2 $\mathrm{km\ s^{-1}}$ and a velocity dispersion of 16 $\pm$ 4 $\mathrm{km\ s^{-1}}$. We display a representative fit, fitting no extra Gauss--Hermite moments (i.e. pure Gaussians) and with 7th order additive and multiplicative Legendre polynomials included in the fitting model, in Figures \ref{fig:pPXF_All} and \ref{fig:pPXF_Hb}. Our second approach is to fit the spectrum, in the same wide ranging \texttt{pPXF} input configurations as used for the \citet{Coelho2014} library, using our observation of the Milky Way GC M3 in the same KCWI setup as a template. After addition of the intrinsic velocity dispersion (5.4 $\mathrm{km\ s^{-1}}$; \citealp{Pryor1993}) and recessional velocity of M3 ($-$141 $\mathrm{km\ s^{-1}}$; \citealp{Smolinski2011}) we again take the medians of the subsequent distributions. This recovers a recessional velocity of 1114 $\pm$ 3 $\mathrm{km\ s^{-1}}$ and a velocity dispersion of 21 $\pm$ 4 $\mathrm{km\ s^{-1}}$, in good agreement with fitting our spectra with the \citet{Coelho2014} library. Additionally, for both approaches the distributions from which the velocity dispersions are quoted have both \texttt{pPXF} fits with h3=h4=0 and fits that have them as free parameters. The quoted errors therefore include any error that may be introduced by non-Guassianity in the absorption features.

We use these two approaches to cover our two largest likely sources of systematic error. Fitting with the \citet{Coelho2014} library has the advantage of the good stellar parameter coverage (-1.3 $\le$ $\mathrm{[Fe/H]}$ $\le$ 0; $\mathrm{[\alpha/Fe]}$ 0 or 0.4), which minimises the possibility of template mismatch, but may include a poorly characterised instrumental resolution. Conversely, fitting with the M3 spectrum has the advantage of the template ideally modelling the instrumental resolution (avoiding our previous assumption of a non-wavelength dependent, Gaussian instrumental line spread function), as it was observed in the same configuration, but assumes a single GC-like template. For the remainder of this work we take the average of these two approaches and combine their uncertainties in quadrature, giving 19 $\pm$ 6 $\mathrm{km\ s^{-1}}$ as our stellar velocity dispersion. For our observations, the centre of VCC 1287 is positioned towards the edge of the KCWI FOV and the long axis of the large slicer is 33''. Noting that the half light radius of VCC 1287 is 41.5'' (circularised), our stellar velocity dispersion represents a flux weighted measurement within $\sim$ 0.8 $\mathrm{R_{e}}$, where larger radii are increasingly under-sampled by our observations. Additionally, in both fits the uncertainties likely capture much of the same information yet we choose to combine these uncertainties in our velocity dispersion to represent an upper limit on the true uncertainty. We include a discussion of the complexities of fitting our KCWI spectroscopic data along with some of the consistency checks performed in Appendix \ref{app:fitting}.

Both of the two recessional velocities extracted by \texttt{pPXF} using different templates are in good agreement with that extracted using our MCMC sky subtraction routine. There is, however, a discrepancy between our calculated recessional velocity and those calculated in \citet{Beasley2016} where the mean recessional velocity for seven compact objects studied was found to be $1071^{+14} _{-15}\ \mathrm{km\ s^{-1}}$. An additional consistency check for our kinematic fitting is provided by a compact object we positioned in our science field. Noting this GC, called GC16 in \citet{Beasley2016}, has a known recessional velocity of $1088 \pm 13\ \mathrm{km\ s^{-1}}$ \citep{Beasley2016} we use it to test our results. To investigate differences between our results and those of \citet{Beasley2016} we extract a spectrum from our science frames by manually taking a 2 by 5 spaxel box centred on GC16 with 5 spaxel columns either side used for background subtraction. We then stack these spectra in the same manner as that of our diffuse stellar light and fit this extracted spectrum with \texttt{pPXF} to determine a recessional velocity. Due to the noticeably lower S/N for GC16 we are only able to constrain the recessional velocity to be of similar magnitude to what we extract for the galaxy. The difference between our recessional velocity and that of \citet{Beasley2016} is $\sim$ 3 pixels in our data. We cannot account for the offset between our recessional velocities and that of \citet{Beasley2016} but, as both are self consistent, we expect this offset to be the result of systematics associated with the use of differing instruments.

Using our calculated recessional velocity we are able to confirm that VCC 1287 is indeed a member of the Virgo cluster with a recessional velocity close to that of the mean recessional velocity of Virgo, i.e. $1144 \pm 18\ \mathrm{km\ s^{-1}}$ \citep{Sandage1990}. Noting that the definition of a UDG has an associated size criterion, that is naturally distance dependent, our confirmation of cluster membership is also a confirmation of VCC 1287 being a UDG.

As measured, our stellar velocity dispersion (19 $\pm$ 6 $\mathrm{km\ s^{-1}}$) lies at mild tension with that of \citet{Beasley2016} who quoted a velocity dispersion from six GCs (excluding their central nucleus candidate, N17) to be $33^{+16}_{-10}\ \mathrm{km\ s^{-1}}$ within 8.1 kpc. We suggest three plausible explanations for this tension. Firstly, we have measured our velocity dispersion using data that samples out to a smaller radius than the \citet{Beasley2016} value. The singular UDG with spatially resolved velocity dispersion measurements (i.e. Dragonfly 44; \citealp{vanDokkum2019b}) displays a rising velocity dispersion profile which may explain the increase in velocity dispersion at larger radii in VCC 1287. Secondly, the GC system may have a larger half-mass radius than the stars while still in equilibrium in the same gravitational potential. If this were the case the GCs are likely to have a higher velocity dispersion at the same radius. Alternatively, we note that the original \citet{Beasley2016} velocity dispersion excluded a GC in their calculation (their N17). N17 was excluded from their velocity dispersion calculation as it was identified as the nucleus, being centrally located. In the new photometry of \citet{Pandya2018} N17 moves away from the photometric centre of the galaxy suggesting it may not be a (now off-centre) nucleus and should be included in calculations. As noted by \citet{Beasley2016}, its inclusion has minimal effect in the resulting calculated recessional velocity and velocity dispersion. Another of their GCs (GC21), however, lies as a $\sim$ 2 standard deviation outlier in the original data set assuming a Gaussian velocity distribution. If we exclude GC21 in calculations, as was done to N17 originally, its recessional velocity becomes > 3 standard deviations from the mean of the remaining GCs, suggesting it may not be in dynamic equilibrium with the rest of the system. Recalculating a simple standard deviation velocity dispersion (excluding GC21), we measure a velocity dispersion of 21 $\mathrm{km\ s^{-1}}$, in better agreement with our stellar velocity dispersion measurement. Given the observational difficulties of obtaining kinematic measurements for UDGs it is frequently unfeasible to obtain more than one estimate for an individual galaxy. It is therefore of paramount importance that we have confidence in these different measurements if we are to sample the mass distribution of UDGs and hence more UDGs with both stellar and GC velocity dispersions are needed.

\subsection{Dynamical Mass}

\begin{figure}
    \centering
    \includegraphics[width = 0.47 \textwidth]{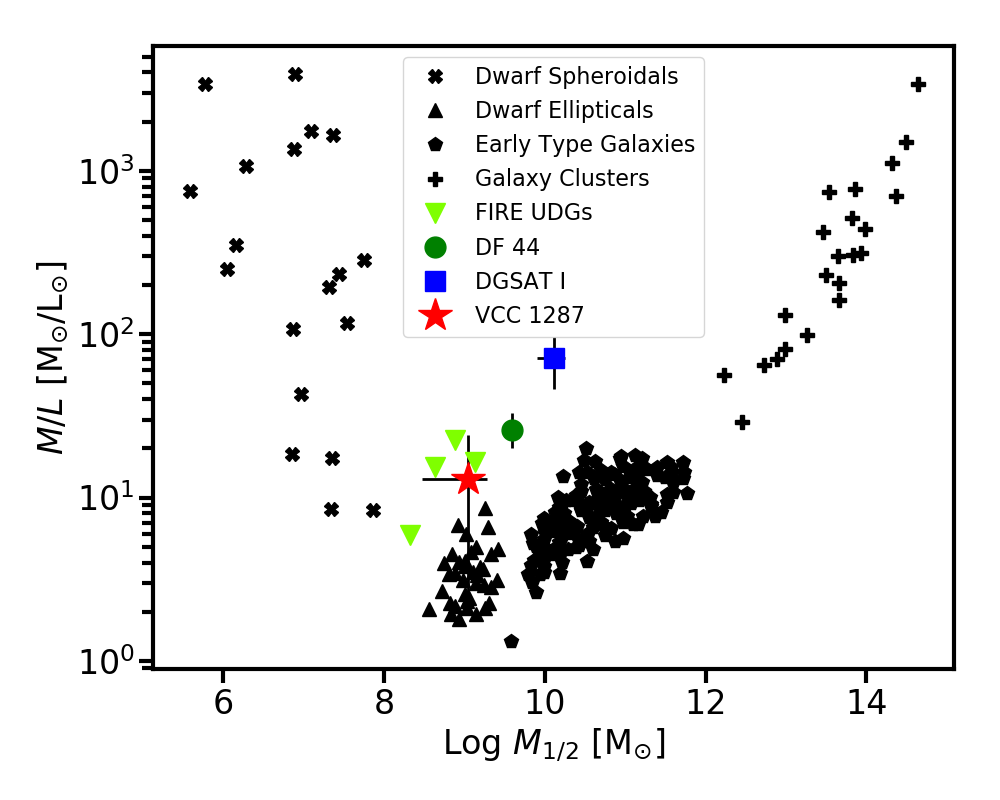}
    \caption{Mass to light ratio vs dynamical mass within half light radius. The mass to light ratio of VCC 1287 (red star -- this work) in addition to the two other UDGs inferred to be dark matter dominated through KCWI stellar kinematics, Dragonfly 44 (green circle -- \citealp{vanDokkum2019b}) and DGSAT I (blue square -- \citealp{Martin-Navarro2019}) for comparison to normal objects. The selection of `normal' objects (black) includes: dwarf spheroidals \citep{Wolf2010} (crosses), dwarf early types \citep{Toloba2014} (triangles), early-type giants \citep{Cappellari2013} (pentagons) and clusters of galaxies \citep{Zaritsky2006} (pluses). A sample of the four model UDGs that meet the \citet{VanDokkum2015} UDG criteria from the FIRE simulations is also included \citep{Chan2018} (inverted green triangles). The observed UDGs reside generally above the classical `U shaped' relation, to a higher mass to light ratio for a given dynamical mass.}
    \label{fig:ML}
\end{figure}

By manipulating the spherical Jeans equation for a pressure supported galaxy, \citet{Wolf2010} derived a simple formula reliant only on observable quantities. For a luminosity-weighted line of sight velocity dispersion ($\sigma$) measured within the 2D projected half light radius ($R_{e}$), the mass within the 3D, deprojected half light\footnote{Deprojected half light radius $\sim \mathrm{\frac{4}{3}}$ projected half light radius \citep{Wolf2010}} radius ($M_{1/2}$) is:

\begin{equation} \label{eqn:wolf}
    M_{1/2}\ =\ 930\ (\frac{\sigma ^{2}}{\mathrm{km^{2} s^{-2}}})\ (\frac{R_{e}}{\mathrm{pc}})\ \mathrm{M_{\odot}}
\end{equation}

As previously mentioned, \citet{Pandya2018} found that the publicly available data used by \citet{Beasley2016} to analyse VCC 1287 had been processed through a pipeline that had over-subtracted the background and hence underestimated both the half light radius and apparent magnitude of the galaxy. Using data reprocessed with a specialised low surface brightness pipeline, \citet{Pandya2018} found an semi-major half light radius of 46.4 arcsec with axis ratio $b/a = 0.8$ (Pandya, private comm.) and an apparent magnitude of $m_{i} = 15.05$. At an assumed Virgo distance of 16.5 Mpc we calculate a projected, circularised $R_{e}$ of 3.3 kpc. The only UDG with a radial profile (i.e. DF44; \citealp{vanDokkum2019b}; \mbox{\citealp{Wasserman2019}}) shows a rising velocity dispersion profile. We choose, however, to assume a flat velocity dispersion profile for VCC 1287 similar to other literature work (e.g. \citealp{Beasley2016, Toloba2018}). Consequently, we do not correct our velocity dispersion due to the field of view of the large slicer being somewhat smaller than the half light radius of the galaxy. Using equation \ref{eqn:wolf} we calculate a mass $M_{1/2}$ of $1.11^{+0.81}_{-0.81} \times 10^{9} \ \mathrm{M_{\odot}}$ for VCC 1287.

In order to get a $M/L$ within this central region, we are also required to re-estimate the luminosity using the updated apparent magnitude from \citet{Pandya2018}. Converting this to an absolute magnitude using our assumed Virgo cluster distance we find an absolute magnitude $M_{i}$ of $-16.05$. Hence we derive a total luminosity of $1.69 \times 10^{8}\ \mathrm{L_{\odot}}$ and a half light radius luminosity of $8.45\pm 0.85 \times 10^{7} \mathrm{L_{\odot}}$ assuming a 10\% error for the photometry. Our calculated $M/L$ is therefore $13^{+11}_{-11}$ within $1\ R_{e}$.

We show a comparison of our new mass measurement along with those of a selection of `normal' objects: dwarf spheroidals \citep{Wolf2010}, early type dwarfs \citep{Toloba2014}, early-type giants \citep{Cappellari2013} and galaxy clusters \citep{Zaritsky2006} in Figure \ref{fig:ML}. In comparison to measurements of dark matter dominated UDGs from KCWI stellar kinematics (\citealp{vanDokkum2019b}; \citealp{Martin-Navarro2019}) we find VCC 1287 is likely not as extremely dark matter dominated. It lies slightly above the scatter in the established `U shaped' relation between $M_{1/2}$ and $M/L$  for normal galaxies to higher $M/L$ for a given $M_{1/2}$.

Strictly speaking the $M/L$ ratios are plotted using slightly different bandpasses in Figure \ref{fig:ML}. While the observed and simulated UDGs are plotted in the $i$/I-bands the `U shaped' relation is plotted in the bluer V-band. This does not effect our conclusions. Due to the red nature of VCC 1287 \citep{Pandya2018} a conversion of our VCC 1287 data to the V-band mildly increases its $M/L$ ratio ($\sim$ 20-30\% depending on the actual colour transformation adopted).

Included in Figure \ref{fig:ML} are the UDGs from the FIRE simulations \citep{Chan2018} with $R_{e} > 1.5$ kpc. We calculate their $M/L$ by converting their $i$-band absolute magnitude to a luminosity in solar units. Despite not creating UDGs as old or extended as those plotted in Figure \ref{fig:ML} some of the FIRE UDGs display $M/L$ above the `U shaped' relation given their dynamical mass. Additionally, while the previously observed UDGs are both of higher dynamical mass and $M/L$ than the FIRE simulated UDGs, our measure is in good agreement with their simulations. Indeed, the FIRE simulations are able to re-create mass profiles similar to the previous mass estimates of VCC 1287 coming from the GCs (see \citealp{Chan2018}; their figure 7). We therefore support the hypothesis, suggested in \citet{vanDokkum2019b}, that the small scatter in this `U shaped' relation for normal galaxies is most likely the result of selection effects as previous studies of this relation did not include UDGs.

We also wish to reinforce the \citet{vanDokkum2019b} conclusion that it is hazardous to interpret Figure \ref{fig:ML} as some UDGs having abnormally large total halo masses for their stellar mass. Due to uncertainties in the shape of the halo mass profile (i.e. core vs cusp) and those in the structural parameters of the halo \citep{Dutton2014} an extension of dynamical mass measurements to total halo properties is non-trivial. In this vein, \citet{vanDokkum2019b} demonstrated a fixed aperture metric for measuring $M/L$ is at least as informative as a measurement of $M/L$ within the half light radius. 

\subsection{VCC 1287 Halo Mass}

\begin{figure}
    \centering
    \includegraphics[width = 0.47 \textwidth]{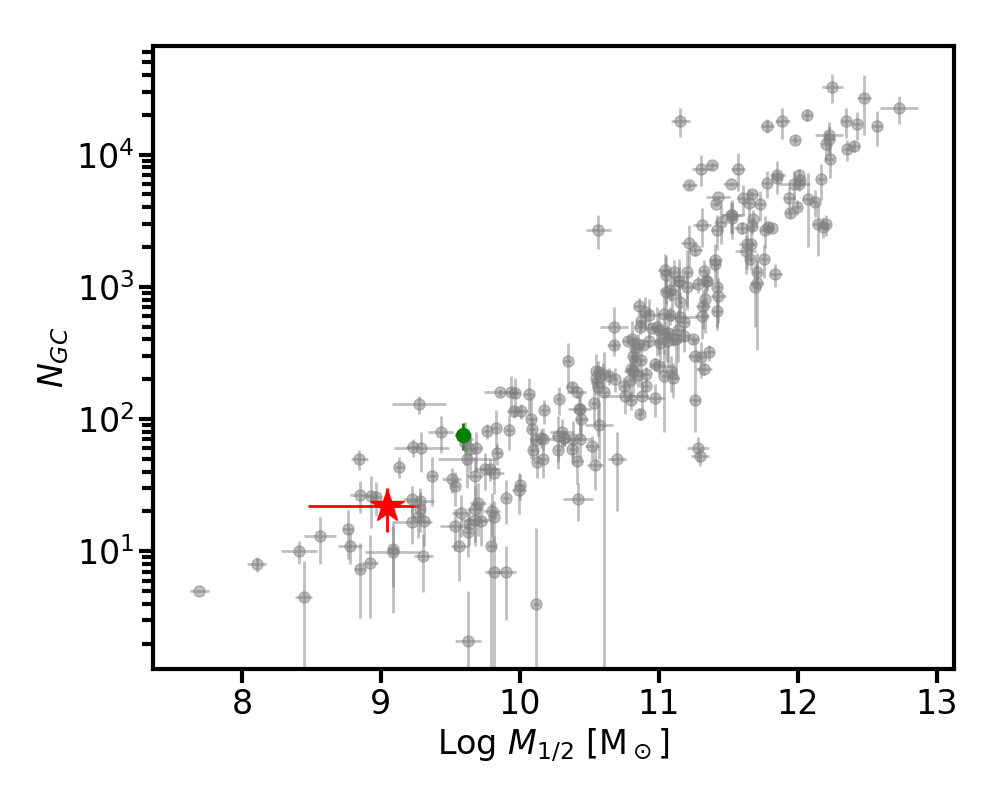}
    \caption{Number of GCs vs dynamical mass within the half light radius. We plot a sample of normal galaxies from \citet{Harris2013} (gray points) along with our measure for VCC 1287 (red star). We include the other UDG with both KCWI stellar kinematics and a GC estimate, Dragonfly 44 (green circle; $N_{GC}$ = 76 $\pm$ 18; \citealp{vanDokkum2017}). VCC 1287 is fully consistent with the $N_{GC} - M_{1/2}$ relation of \citet{Harris2013}, lending support to the idea it will also obey the $N_{GC} - M_{200}$ relation.}
    \label{fig:harris}
\end{figure}

While the extrapolation of a measured dynamical mass into a total halo mass requires the non-trivial assumption of a halo profile, estimating a galaxy's halo mass from its GC system has a well established relation \citep{Spitler2009, Harris2017, Burkert2019}. In Figure \ref{fig:harris} we plot a sample of normal galaxies from \citet{Harris2013} with measurements of both $M_{1/2}$ and $N_{GC}$ along with our measurement for VCC 1287. We include the only other overmassive UDG with both KCWI stellar kinematics and an accurate GC estimate, Dragonfly 44 ($N_{GC}$ = 76 $\pm$ 18; \citealp{vanDokkum2017}). Similar figures have also been made of UDGs with non-KCWI dynamics (see \citealp{Toloba2018} figure 4). VCC 1287 is fully consistent with the $N_{GC} - M_{1/2}$ relation of \citet{Harris2013} which suggests it should also obey the $N_{GC} - M_{200}$ relation. 

Additionally, Figure \ref{fig:harris} demonstrates the agreement between the other UDG plotted, Dragonfly 44, and the $N_{GC} - M_{1/2}$ relation. Currently this UDG is the only one with an independent total halo mass measurement aside from that coming from its GC system \citep{vanDokkum2019b, Wasserman2019}. Both the independent total halo mass measurement, and that coming from the GC system, are similar. Again, we suggest it is therefore likely that VCC 1287 will also obey similar relations for its total halo mass. As we have reason to believe that the GC system counts of VCC 1287 may be able to provide an accurate halo mass estimate we can now seek to use our dynamical mass measure to infer general properties about the halo in which the galaxy may lie.

We use the \citet{Burkert2019} relation ($M_{200} = 5 \times 10^{9} \mathrm{M_{\odot}}\ N_{GC} $) in conjunction with the number of GCs ($N_{GC} = 22 \pm 8$; \citealp{Beasley2016}) to calculate $M_{200} = 1.1 \pm 0.4 \times 10^{11} \mathrm{M_{\odot}}$ for the total halo mass of VCC 1287. VCC 1287 now has three estimates of mass available at different radii: our stellar velocity dispersion, a tracer mass estimate from GC velocities, and a halo mass from the GC numbers.

\begin{figure}
    \centering
    \includegraphics[width = 0.47 \textwidth]{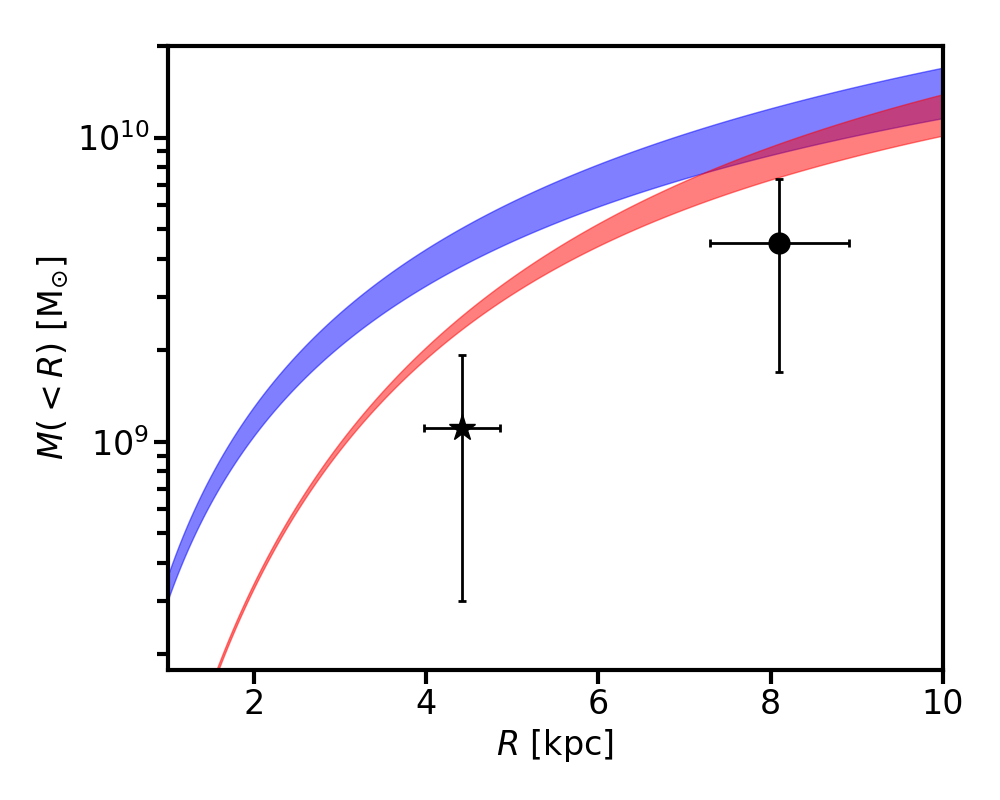}
    \caption{A comparison of the halo mass estimate from GCs and the dynamical mass estimates for VCC 1287. We plot the \citet{Beasley2016} GC tracer mass estimate (circle) and our stellar velocity dispersion dynamical mass measurement (star). A total halo mass estimate based on the GC number -- halo mass relation of \citet{Burkert2019} is plotted using two differing assumptions of halo profile (blue region -- NFW cusp profile; red region -- \citealp{DiCintio2014} halo core profile). For both dynamical mass measurements to be consistent with the halo mass estimate from the number of GCs ($N_{GC} = 22 \pm 8$; \citealp{Beasley2016}) either a lower halo concentration parameter or the presence of a dark matter core is required.}
    \label{fig:estimators}
\end{figure}

In Figure \ref{fig:estimators} we plot the halo mass inferred from GC counting along with our dynamical mass estimate and the \citet{Beasley2016} tracer mass estimate from GC motions. To plot the halo mass inferred using the \citet{Burkert2019} relation we assume two differing halo profiles - analytically derived NFW  profiles \citep{Navarro1996} and empirically derived \citet{DiCintio2014} profiles. These halo profiles are plotted in accordance with the method outlined in the appendix of \citet{DiCintio2014} with the total halo mass being enclosed at 200 times the critical density of the Universe and the concentration parameter being set using equation 8 of \citet{Dutton2014}. When using this method for NFW profiles we do not correct the concentration parameter from that of \citet{Dutton2014} as is done for \citet{DiCintio2014} profiles. 

\begin{figure*}
    \centering
    \includegraphics[width = 0.9 \textwidth]{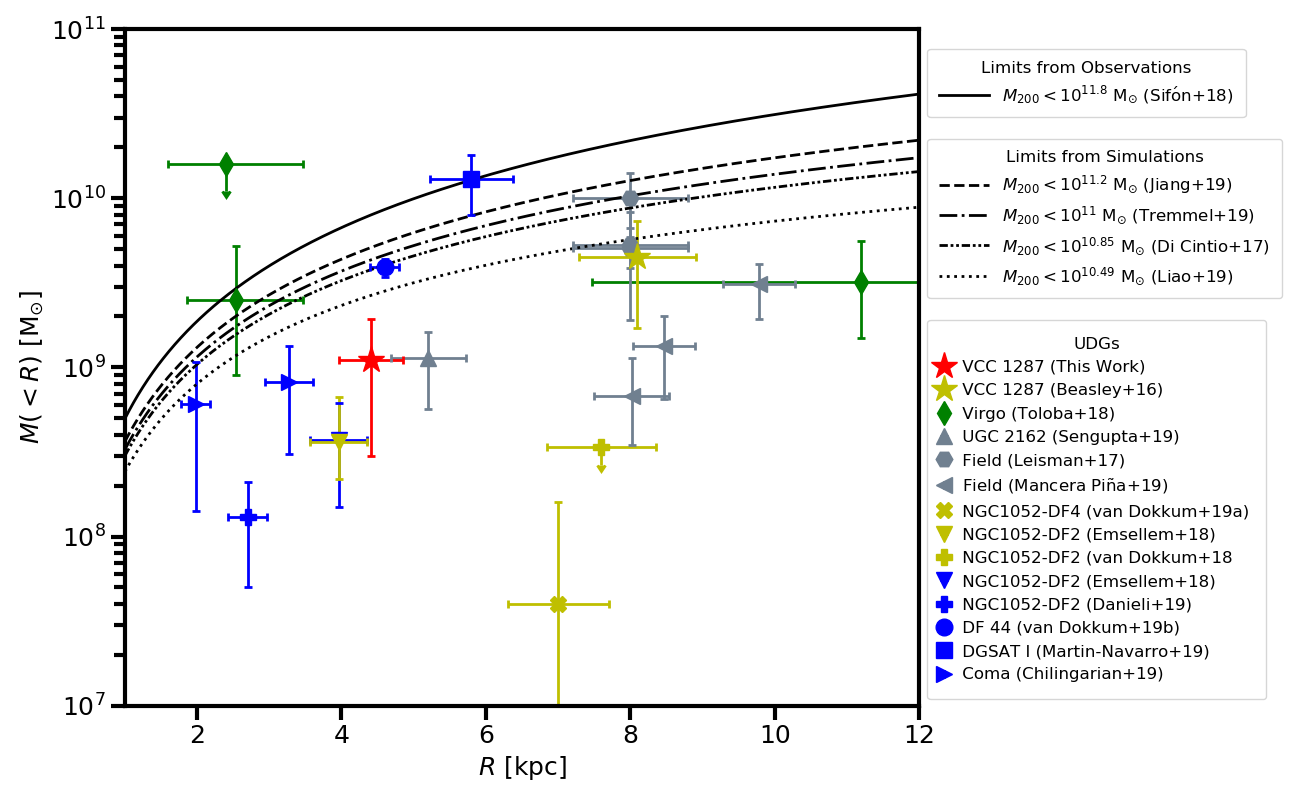}
    \caption{A comparison of measured UDG masses with upper limits from observations and theory. Mass measurements from the literature are included for:
    VCC 1287 (yellow star -- \citealp{Beasley2016}), \citet{Toloba2018} UDGs (diamonds), UGC 2162 (triangle -- \citealp{Sengupta2019}), \citet{Leisman2017} UDGs (pentagons),  NGC1052--DF4 (cross -- \citealp{vanDokkum2019}), NGC1052--DF2 (inverted triangles -- \citealp{Emsellem2018} and pluses -- \citealp{Danieli2019, vanDokkum2018}), Coma UDGs (right triangles -- \citealp{Chilingarian2019}$^{9}$), Field UDGs (left triangles -- \citealp{ManceraPina2019}$^8$), Dragonfly 44 (circle -- \citealp{vanDokkum2019b}) and DGSAT I (square -- \citealp{Martin-Navarro2019}). Observational measurements are colour coded by the method of mass estimation: HI dynamics (gray), GC velocity dispersions (green), tracer mass estimation (yellow) and stellar velocity dispersions (blue). We also plot our stellar velocity dispersion measurement for VCC 1287 (red star). UDG NFW profiles are plotted for upper halo mass limits from: weak lensing observations (solid line -- \citealp{Sifon2018}), evolved NIHAO simulated galaxies (dashed line -- \citealp{Jiang2019}, dashed--dot--dotted line -- \citealp{DiCintio2017}), ROMULUS simulated galaxies (dash--dotted line -- \citealp{Tremmel2019}) and the Auriga simulations (dotted line -- \citealp{Liao2019}). NFW profiles represent upper limits for similar mass halos that have cores or lower concentration parameters as may be expected for UDGs \citep{DiCintio2017, Jiang2019, Carleton2018}. To date observations are consistent with upper limits from \citet{DiCintio2017}, \citet{Sifon2018}, \citet{Jiang2019} and \citet{Tremmel2019} although the assumption that these upper limits can be represented as NFW profiles is likely over-generous . 
    }
    \label{fig:compare}
\end{figure*}

Theoretical efforts to simulate the formation of UDGs frequently rely either on the creation of a cored dark matter profile (e.g. \citealp{DiCintio2017}) or on its pre-existence (e.g. \citealp{Carleton2018}). \citet{Jiang2019} also found UDGs in the NIHAO simulation to lie in cored dark matter profiles along with having concentration parameters lower than that expected based on dwarf galaxy and $L^{*}$ galaxy control samples. It is yet to be established whether these low concentration parameters are the cause of, or result of, UDG formation \citep{Jiang2019}. In both the case of a cored halo or one with a lower concentration parameter the enclosed mass of the profile would be biased towards lower values causing NFW profiles to represent an upper limit to the enclosed mass. For this reason we also choose to include the \citet{DiCintio2014} halo profile as these models naturally produce cores which may be more representative of the actual profile of UDGs.

There exists a slight tension between the dynamical masses and the halo mass, as plotted, calculated with the \citet{Burkert2019} relation. This tension may be the result of incorrect assumptions made in converting observations into dynamical masses or, more likely, incorrect halo profile assumptions made in plotting for the full halo mass inferred from GC counts (e.g. VCC 1287 may have a lower concentration parameter than we calculate with \mbox{\citealp{Dutton2014}}). In order for the dynamical mass measurements to be consistent with the halo mass estimate from the \mbox{\citet{Burkert2019}} relation the dark matter halo of VCC 1287 must either be cored (e.g. the \citealp{DiCintio2014} profile of Figure \ref{fig:estimators}) or have a low concentration parameter.

We note that the total halo mass from the \citet{Burkert2019} relation implies a halo more massive that what is expected from the stellar mass - halo mass relation similar to other UDGs in the Coma Cluster \citep{Forbes2020}. Indeed, both the stellar mass and GC system richness of VCC 1287 is similar to these Coma Cluster UDGs which have been dubbed as `failed' galaxies (e.g. \citealp{Forbes2020}). On a similar line of argument \citet{Beasley2016} suggest it likely that VCC 1287 is also a `failed' galaxy UDG.

Using the Illustris-dark simulations, \citet{Carleton2018} made predictions for the stellar velocity dispersion within the half light radius of tidally formed UDGs in predominately cored dark matter halos. They predicted an average line of sight velocity dispersion of $14\ \mathrm{km\ s^{-1}}$, with the 10th -- 90th percentile range existing between 9 and 23 $\mathrm{km\ s^{-1}}$, in good agreement with our measure for VCC 1287. In similar analysis done by \citet{Sales2019} using the IllustrisTNG simulations (where baryonic effects are modelled, in contrast to the \citealp{Carleton2018} dark matter only version of Illustris), UDGs with velocity dispersions closer to that of the regular dwarf population, for their stellar mass, are consistent with being ``tidal UDGs'' - where tidal effects in the cluster are important for the creation of UDGs. Given the agreement of our stellar velocity dispersion with both of the Illustris simulations and the FIRE simulations of \citet{Chan2018} a formation scenario involving a combination of strong stellar feedback and/or tidal effects seems plausible for VCC 1287. In order to further test these hypotheses we will need to conduct stellar population analysis to measure the alpha element enhancement known to be prevalent in galaxies with strong stellar feedback. Unfortunately this is beyond the scope of this work as the spectral region ($\mathrm{Mgb}$) to measure such enhancement is not used.

\subsection{UDG Halo Masses} \label{sec:halomasses}

Theoretical studies of UDG formation have predominantly predicted them to reside in $M_{200} \sim 10^{10} - 10^{11} \mathrm{M_{\odot}}$ halos \mbox{\citep{Amorisco2016}} with upper limits to their halo mass distribution at $M_{200} = 10^{10.49} \mathrm{M_{\odot}}$ \citep{Liao2019}, $M_{200} = 10^{10.85} \mathrm{M_{\odot}}$ \mbox{\citep{DiCintio2017}}, $M_{200} = 10^{11} \mathrm{M_{\odot}}$ \citep{Tremmel2019} and $M_{200} = 10^{11.2} \mathrm{M_{\odot}}$ \citep{Jiang2019}\footnote{Despite both \citet{DiCintio2017} and \citet{Jiang2019} beginning with a similar sample of galaxies from the NIHAO simulations \citet{Jiang2019} evolved their galaxies using the updated version the smoothed particle hydrodynamics code GASOLINE. This, combined with the addition of group UDGs to the studied sample, results in a slightly different upper limit to \citet{DiCintio2017}.}. Weak lensing observations place a 95\% confidence interval upper limit for the mass of UDGs at $M_{200} = 10^{11.8} \mathrm{M_{\odot}}$. To allow comparison to observations in Figure \ref{fig:compare} we plot these upper limits as NFW profiles which, as previously mentioned, are themselves upper limits to the halo mass profiles expected for UDGs due to the possible presence of cores (\mbox{\citealp{DiCintio2017}}, \mbox{\citealp{Jiang2019}}, \mbox{\citealp{Carleton2018}}) and lower concentration parameters \citep{Jiang2019}. Included in Figure \ref{fig:compare} are a variety of UDGs with dynamical masses inferred from HI dynamics (\citealp{Leisman2017}; \citealp{Sengupta2019}; \mbox{\citealp{ManceraPina2019}})\footnote{We only plot \citet{ManceraPina2019} UDGs that meet our definition and are not already included in the \citet{Leisman2017} sample plotted.}, TME from GCs and/or planetary nebulae (\citealp{Beasley2016}; \citealp{vanDokkum2018}; \mbox{\citealp{Emsellem2018}}; \citealp{vanDokkum2019}), GC velocity dispersions \citep{Toloba2018}) and stellar velocity dispersions (\citealp{Danieli2019}; \citealp{vanDokkum2019b}; \citealp{Emsellem2018};  \citealp{Martin-Navarro2019}; \citealp{Chilingarian2019})\footnote{Here we only plot \citet{Chilingarian2019} UDGs that meet our definition.}. Due to the consistency between our stellar velocity dispersion measure and the GC velocity dispersion measure of \citet{Beasley2016} we do not plot their enclosed mass in Figure \ref{fig:compare} for ease of viewing. We note that many of the objects plotted were observed due to their abnormally large sizes (even for UDGs) or abnormally large GC populations (a known indicator of high halo mass \citep{Burkert2019}) making them likely unrepresentative of the overall UDG mass distribution. 

Based on Figure \ref{fig:compare} we suggest that the observational upper limit of \citet{Sifon2018} is likely representative of that of the UDG halo mass distribution, suggesting that few UDGs should have halo masses > $10^{11.8} \mathrm{M_{\odot}}$. Conversely, we find that some of the data are in conflict with the theoretical upper limits from the Auriga simulations \citep{Liao2019}. Although some of these observations are of UDGs in denser environments, which are not simulated in \citet{Liao2019}, `puffy dwarf' UDGs are not completely representative of current observations. Testing predictions from the simulations of \citet{DiCintio2017}, \citet{Jiang2019} and \citet{Tremmel2019} is more complex. Some of the data are not in conflict with theoretical upper limits from the those simulations, as plotted, but this does not indicate agreement between these upper limits and observations. Both \mbox{\citet{DiCintio2017}} and \citet{Jiang2019} predict cored dark matter profiles while \citet{Tremmel2019} is unable to resolve dark matter cores in their simulated UDGs. This may mean current observations are in tension with \citet{DiCintio2017}, \mbox{\citet{Jiang2019}} and \mbox{\citet{Tremmel2019}} predicted upper limits also, due to the plotting of their upper limits as NFW profiles in Figure \ref{fig:compare}.

Additionally, with the revision of a lower halo mass for Dragonfly 44 in \citet{vanDokkum2019b}, there exists dwindling evidence for UDGs residing in $L^{*}$-like halos. We suggest the current use of the term `failed $L^{*}$ galaxy' for some UDGs to be misleading and the term `failed galaxy' is a more accurate representation of the current state of observations for these objects (see also similar ideas in \citealp{Peng2016}; \citealp{Beasley2016b}).

\section{Conclusions}
In this work we have presented a modern spectroscopic PCA method for accurate sky subtraction using offset sky frames for KCWI. We use this method to extract the low surface brightness light from the UDG VCC 1287 in order to analyse its kinematic properties and test the consistency of the multiple mass estimates available. Our main conclusions are as follows:

\begin{itemize}
    \item We measure a stellar recessional velocity of $1116 \pm 2\ \mathrm{km\ s^{-1}}$ for VCC 1287 confirming its association with both the GCs in \citet{Beasley2016} and with the Virgo cluster. Association with the Virgo cluster formalises its large size ($R_{e} = 3.3$ kpc) and thus its status as a UDG.
\end{itemize}
\begin{itemize}
    \item We measure a stellar velocity dispersion of $19 \pm 6\ \mathrm{km\ s^{-1}}$ for VCC 1287 implying a dynamical mass $M_{1/2}$ of $1.11^{+0.81}_{-0.81} \times 10^{9} \ \mathrm{M_{\odot}}$ within 4.4 kpc. From our dynamical mass we calculate a $M/L$ ratio of $13^{+11}_{-11}$ within the half light radius.We find VCC 1287 likely lies towards the upper end of the scatter in the established `U shaped' relation for normal galaxies - it sits at a higher $M/L$ ratio for a given measured dynamical mass. Using this, combined with other UDG measurements, we support the conclusion that the small scatter in this relation is at least partially related to the previous non-inclusion of low surface brightness galaxies.
\end{itemize}
\begin{itemize}
    \item Our dynamical mass, along with the number of GCs associated with VCC 1287, are consistent with the \citet{Harris2013} $N_{GC}\ -\ M_{1/2}$ relation, suggesting we can estimate accurate halo masses from the GC system counts. In order for our dynamical mass measurement, along with the mass estimate from the GC motions \citep{Beasley2016}, to agree with the halo mass estimate coming from GC counts, VCC 1287 must reside in either a cored halo or one with a lower concentration parameter. Additionally, we suggest that the agreement between our data and predictions from the simulations of \citet{Chan2018}, \citet{Carleton2018} and \citet{Sales2019} makes it plausible that tidal effects and/or stellar feedback played a key role in the formation of VCC 1287. A stellar population analysis could test the feedback hypothesis as galaxies with strong stellar feedback are known to exhibit alpha element enhancement.
\end{itemize}

\section*{Acknowledgements}

We thank the anonymous referee for reading the paper carefully and providing thoughtful comments which allowed the refinement and enhancement of this work. We thank A. Alabi for sharing his knowledge of kinematic fitting and for assisting in the observations for this work. JSG wishes to thank E. N. Taylor, M. Sinha, A. Wasserman, M. Beasley and M. Durr\'e for insightful discussions that helped improve this work. The data presented herein were obtained at the W. M. Keck Observatory, which is operated as a scientific partnership among the California Institute of Technology, the University of California and the National Aeronautics and Space Administration. The Observatory was made possible by the generous financial support of the W. M. Keck Foundation. The authors wish to recognise and acknowledge the very significant cultural role and reverence that the summit of Maunakea has always had within the indigenous Hawaiian community.  We are most fortunate to have the opportunity to conduct observations from this mountain. We thank the staff of the W. M. Keck Observatory for their assistance in using the telescope and wish to express particular gratitude to Luca Rizzi in helping us fix issues that presented themselves. We have also made heavy use of a number of open source packages in the creation of this work. We therefore wish to thank contributors to \textit{Astropy} \citep{astropy2018}, \textit{Scipy} \citep{Scipy}, \textit{scikit--learn} \citep{scikit-learn}, \textit{Numpy} \citep{numpy}, \textit{Seaborn} \citep{seaborn}, \textit{corner} \citep{corner} and \textit{emcee} \citep{EMCEE} for making their software publicly available. JSG acknowledges financial support received through a Swinburne University Postgraduate Research Award throughout the creation of this work. AFM has received financial support through the Post-doctoral Junior Leader Fellowship Programme from ``La Caixa'' Banking Foundation (LCF/BQ/LI18/11630007). AJR was supported by National Science Foundation grant AST-1616710, and as a Research Corporation for Science Advancement Cottrell Scholar. JPB gratefully acknowledges support from National Science foundation grants AST- 1518294 and AST-1616598.



\bibliographystyle{mnras}
\bibliography{bibliography.bib} 



\appendix
\section{KCWI Flat Fielding} \label{app:flat}

\begin{figure}
    \centering
    \subfigure[]{
    \includegraphics[width = .21\textwidth]{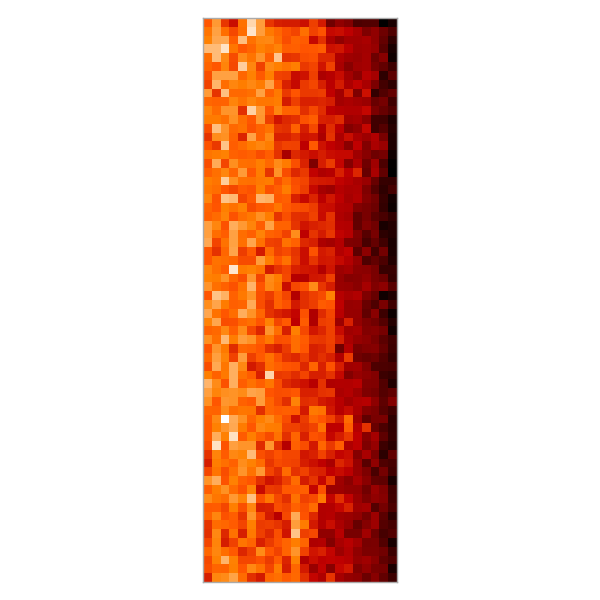}
    }
    \qquad
    \subfigure[]{
    \includegraphics[width = .21\textwidth]{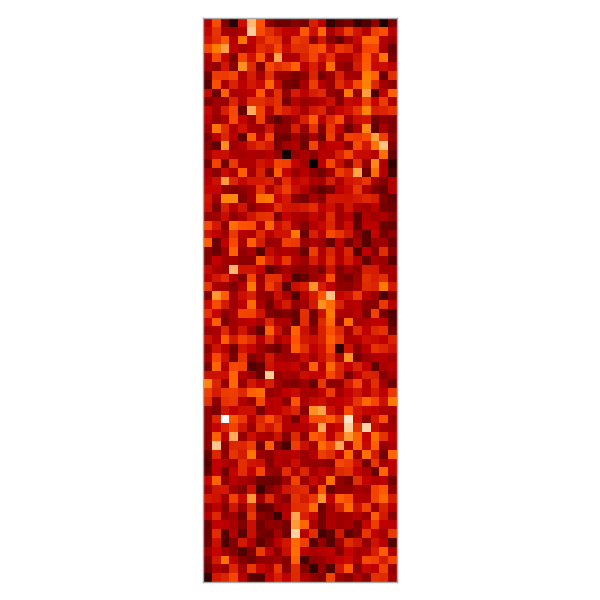}
    }\\
    \subfigure[]{
    \includegraphics[width = .45\textwidth]{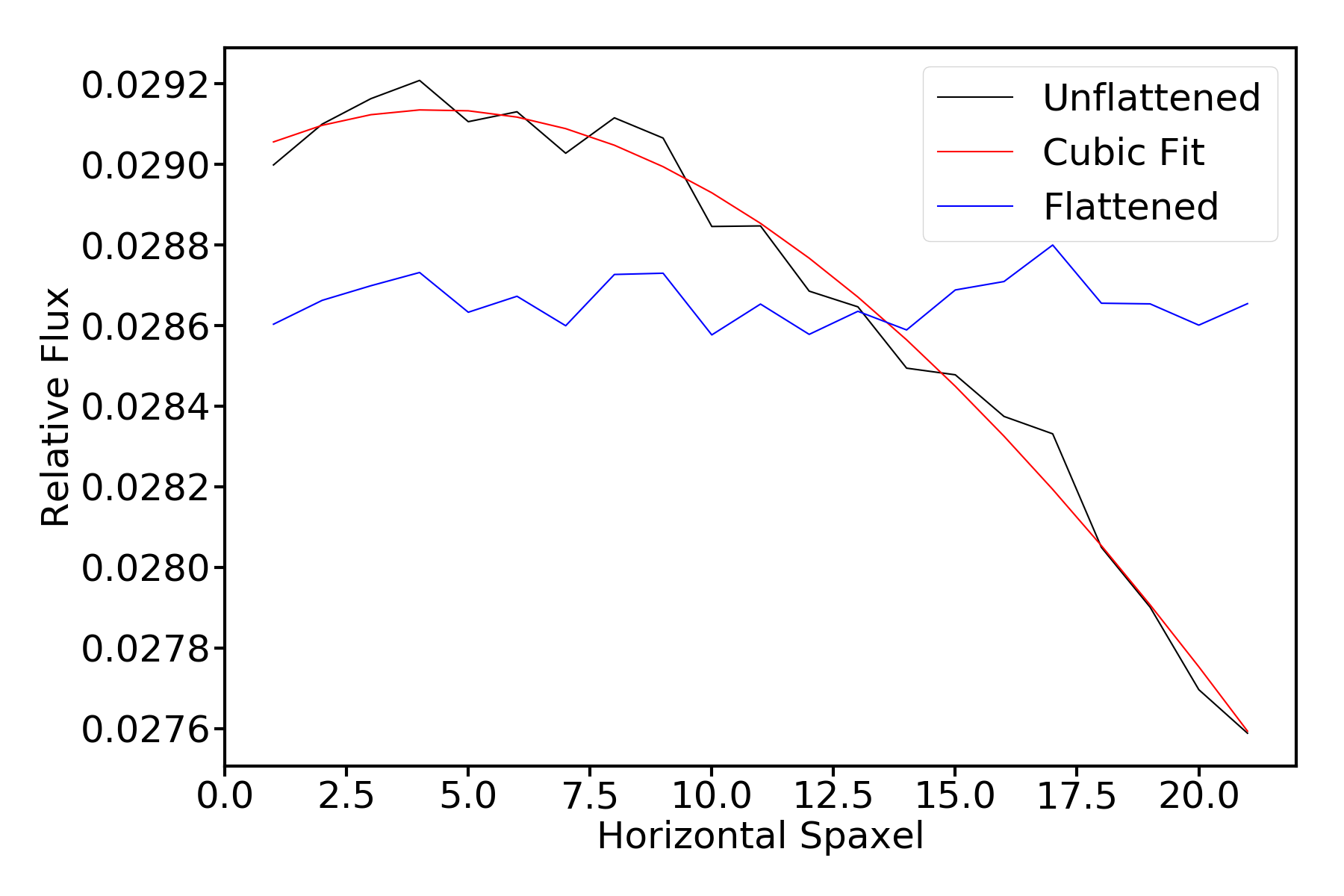}
    }
    
    \caption{A visual representation as to how we correct the gradient discovered in our data. \textit{a)} A 2d visualisation of a data cube that has been processed by the KCWI pipeline. \textit{b)} The same cube after we have applied our flat fielding fix. \textit{c)} An example wavelength slice from the cube (black) where the unflattened data are fitted with a cubic function (red) and flattened. We show the result of this flattening in blue.}
    \label{fig:flatfix}
\end{figure}

\begin{figure}
    \centering
    \includegraphics[width = 0.45\textwidth]{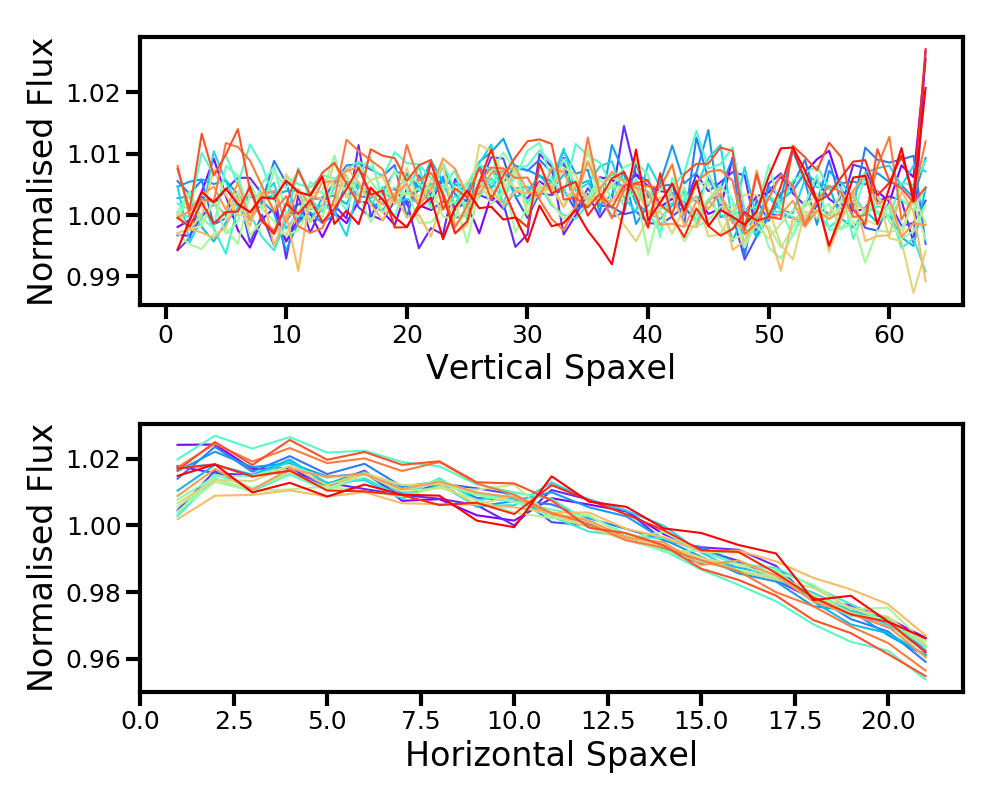}
    \caption{Collapsed vertical and horizontal slices in our data cubes. \textit{Upper Panel}: The result of collapsing a data cube along its horizontal and wavelength directions to create a single median vertical slice through the cube. \textit{Lower Panel}: The same but we collapse the cube vertically to create horizontal slices. Lines of different colours denote differing data cubes from the VCC 1287 observing runs. We note a clear horizontal gradient in all of our frames.}
    \label{fig:flat}
\end{figure}

Immediately apparent upon visual inspection of our science frames after reduction through the KDERP pipeline was a horizontal gradient across the field of view (see Figure \ref{fig:flatfix} \textit{a)} ). We originally thought this gradient may simply be that of the target galaxy but further investigation of our sky exposures along with twilight flats saw a similar effect. We also checked our own archival data targeting differing objects on differing observing runs, finding the gradient present in all cases. In order to try to quantify the error in the flat fielding we collapse our 20 science and sky frames into single vertical and horizontal slices as is shown in Figure \ref{fig:flat}. We note that vertically our frames are flat at around the $\sim$ 1\% level. Horizontally however, we see a clear $\sim$ 6 \% gradient that seems to vary between about $\sim$ 4 \% to $\sim$ 8 \% on a frame to frame basis. Due to this variation we have chosen to correct our images individually by writing our own simple python script to be applied post \textit{KDERP} pipeline.

This python script reads in the output `ocubes', that have been reduced by the KDERP pipeline, and then crops them to both the good wavelength and good spatial range. We then select each wavelength slice of the cube individually and median collapse the cube into a single horizontal spaxel slice. We choose median collapsing as it will be less sensitive to bright objects in the frame such as the GC visible in our VCC 1287 exposures. After building this model for the background in the image we then fit a simple cubic polynomial to our data and divide through to flatten the wavelength slice. We then loop through all wavelength slices in our data cube. We show the application of this code in Figure \ref{fig:flatfix}. Once we have applied our post-pipeline correction we find our data cubes to be flat horizontally at about the $\sim$ 1 \% level. We note this code is unable to function in a case of a significantly non-smooth background such as that of our M3 GC observations. Here we have such large signal to noise in a single exposure so as to minimise the need for this correction. An unfortunate side effect of this code will be the removal of any gradient that may exist in the UDG. Its measurement was not vital for our science and, as we are using KCWI in light-bucket mode, we collapse the data cube over any gradient that may exist. Keck staff are aware of and investigating this issue. 

\section{Fitting Velocity Dispersions} \label{app:fitting}
\begin{figure}
    \centering
    \subfigure[]{
    \includegraphics[width = .45\textwidth]{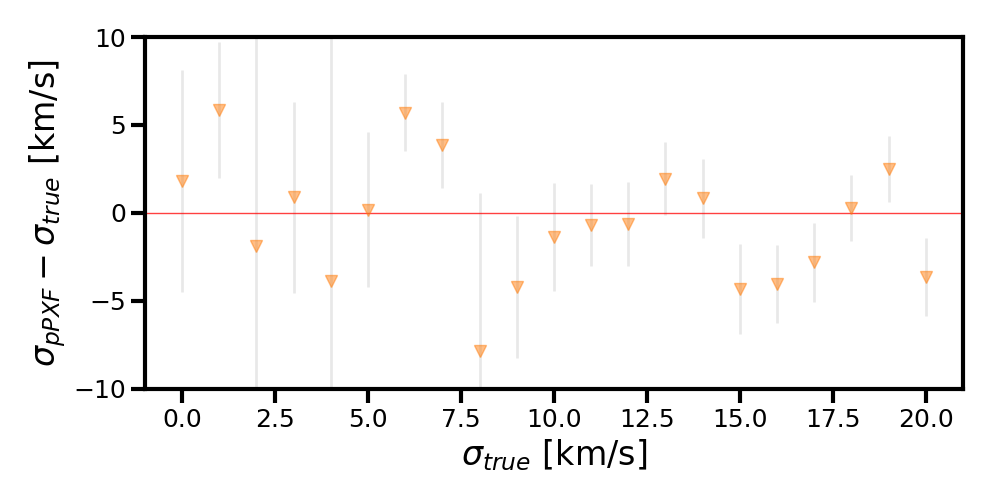}
    }\\
    \subfigure[]{
    \includegraphics[width = .45\textwidth]{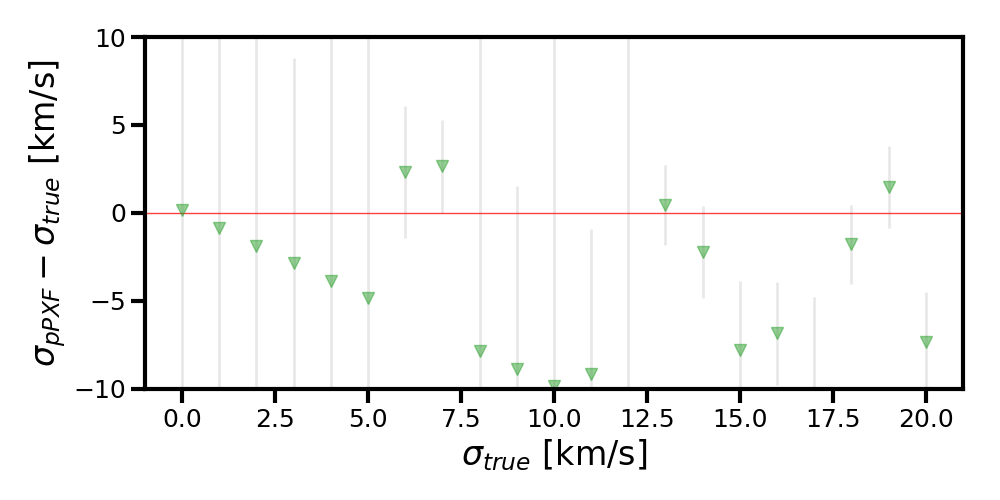}
    }
    
    \caption{The results of our tests fitting mock spectra using \texttt{pPXF} in the low S/N (here we show a S/N of 15), low velocity dispersion regime. We plot the error in the extracted \texttt{pPXF} velocity dispersion ($\sigma_{pPXF}$) against the velocity dispersion inserted into the data ($\sigma_{true}$). \textit{Upper:} The results of fitting the mock data using the \citet{Coelho2014} library. \textit{Lower:} The results of fitting the mock data using the ELODIE library. Low S/N and a low velocity dispersion do not limit our ability to recover a velocity dispersion when using the \citet{Coelho2014} library, while the ELODIE library gives systematically low velocity dispersions.}
    \label{fig:SNtest}
\end{figure}

\begin{figure*}
    \centering
    \includegraphics[width = 0.9\textwidth]{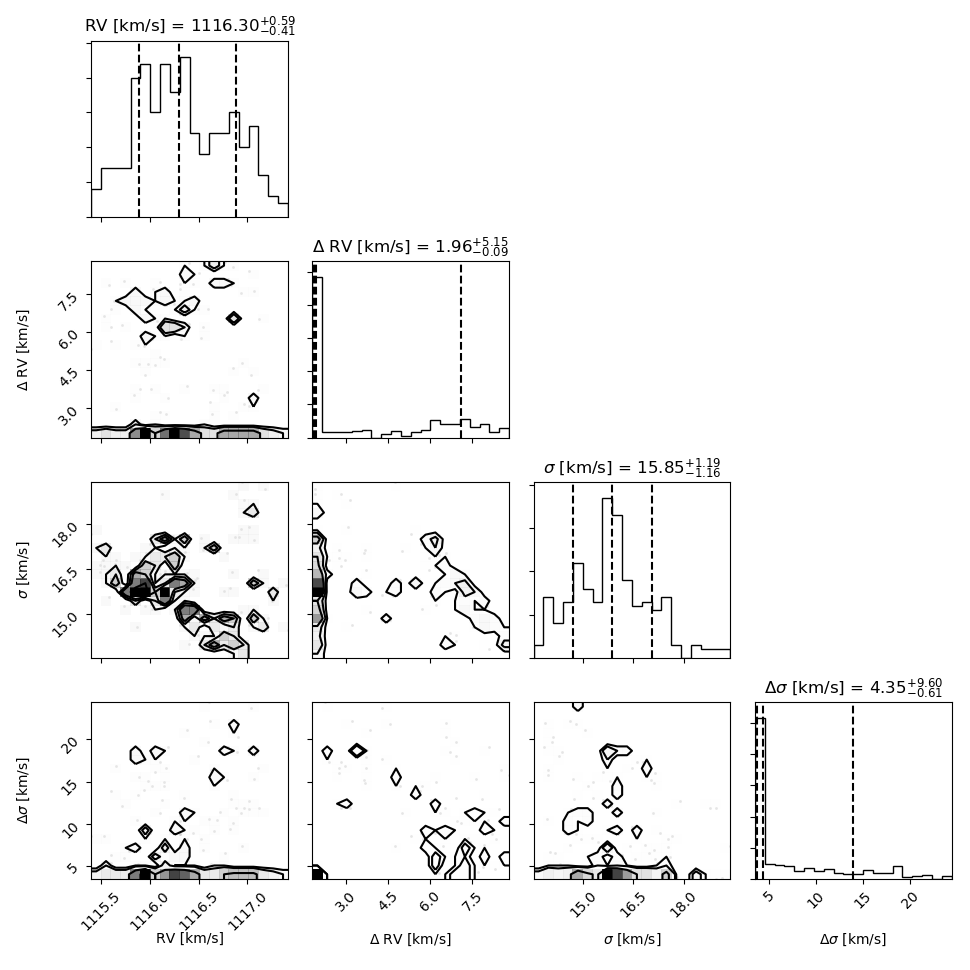}
    \caption{The results for recessional velocity (RV) and velocity dispersion ($\sigma$) with associated errors ($\Delta$ RV; $\Delta \sigma$) of fitting our VCC 1287 data using 242 different parameter combinations of \texttt{pPXF} using the \citet{Coelho2014} stellar library. Here we have removed the < 2\% of fits that we deem too imprecise to have fit the data well (i.e. those with errors > 25 $\mathrm{km\ s^{-1}}$ in recessional velocity and/or velocity dispersion). We find a good convergence not only to a singular recessional velocity and velocity dispersion but also to a similar value for their errors. We take the median values from the subsequent parameter distributions and use them in the main body of this work (i.e. recessional velocity - 1116 $\pm$ 2 $\mathrm{km\ s^{-1}}$; velocity dispersion - 16 $\pm$ 4 $\mathrm{km\ s^{-1}}$). Please note that this is not MCMC, here we simply display the output of our test in a similar manner to what is standard for MCMC testing.}
    \label{fig:corner}
\end{figure*}

\begin{figure*}
    \centering
    \includegraphics[width = 0.9\textwidth]{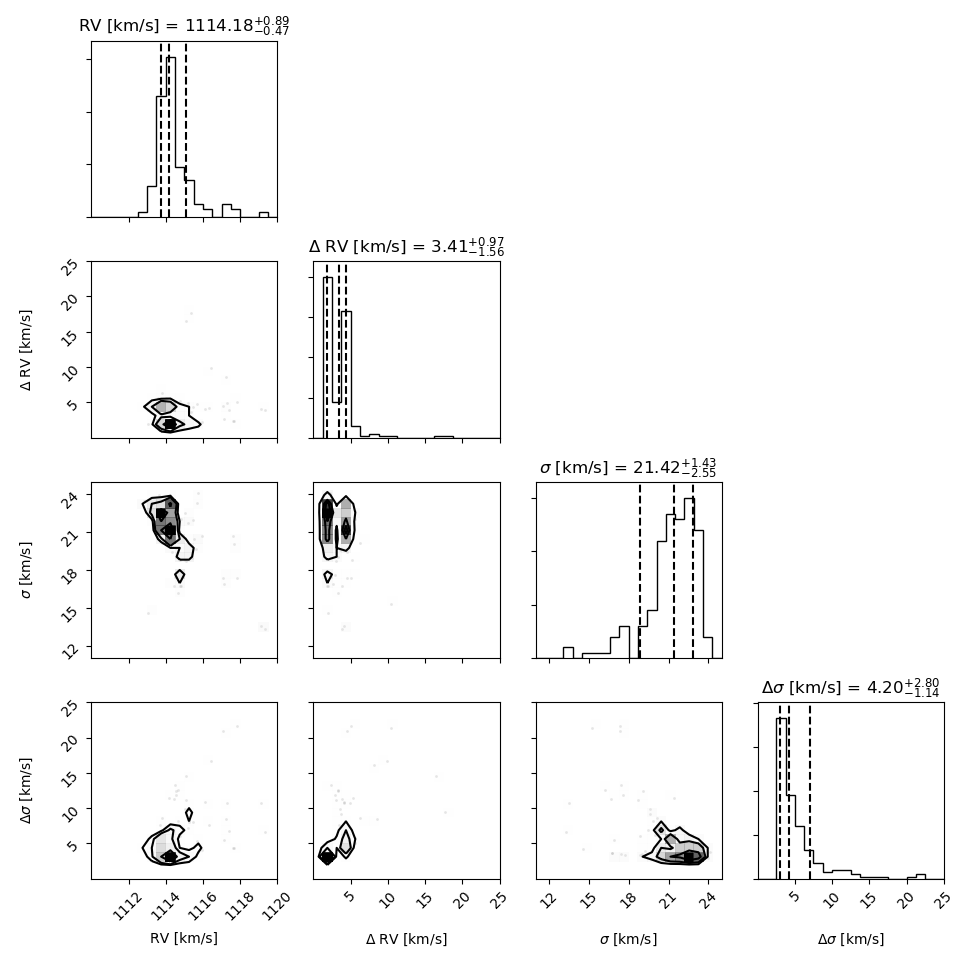}
    \caption{The same as Figure \ref{fig:corner} but using our observation of the Milky Way GC M3 as the \texttt{pPXF} template instead. We have added the intrinsic velocity dispersion (5.4 $\mathrm{km\ s^{-1}}$; \citealp{Pryor1993}) and recessional velocity of M3 (-141 $\mathrm{km\ s^{-1}}$; \citealp{Smolinski2011}) from these fits prior to plotting. Again, we take the median values from the subsequent parameter distributions and use them in the main body of this work (i.e. recessional velocity - 1114 $\pm$ 3 $\mathrm{km\ s^{-1}}$; velocity dispersion - 21 $\pm$ 4 $\mathrm{km\ s^{-1}}$).}
    \label{fig:cornerm3}
\end{figure*}

\subsection{Sky Subtraction}

Before examining any systematic effects we first wish to rule out any errors that may be introduced by our PCA sky subtraction routine described in Section \ref{sec:skysub}. In order to do this we performed an expanded version of the test described in Section \ref{sec:fam}. We take each of our 10 sky frames and insert mock galaxy flux into them before sky subtracting this spectrum using the remaining 9 sky frames, stacking the data and extracting kinematics. We smooth the inserted templates to 0/15/33 $\mathrm{km\ s^{-1}}$ above the instrumental velocity dispersion of KCWI in our configuration (i.e. 25 $\mathrm{km\ s^{-1}}$). In all three cases we are able to recover both the recessional velocity and velocity dispersion of the template we have inserted into the data (in the case where the template is smoothed to the instrumental resolution we instead recover an upper limit for the velocity dispersion). We conclude our sky subtraction routine does not affect the recessional velocity or velocity dispersion of the spectrum it recovers.

\subsection{Systematic Considerations}

In the case of UDGs, velocity dispersions will frequently be around or below the instrumental resolution, the data will frequently have low S/N and the UDG may harbour an atypical stellar population (e.g. DGSAT I; \citealp{Martin-Navarro2019}). Thus, there are many systematic sources of error that must be considered when fitting for a velocity dispersion. Here we briefly discuss the effect that three of them: low spectral resolution (both for the templates and the instrument itself), low S/N data and template mismatch, has on our ability to recover a velocity dispersion.

Here we seek to test the effects of the low S/N and low velocity dispersions we expect for UDGs. To do this we select one template each from the \citet{Coelho2014} and ELODIE \citep{ELODIE2004} stellar libraries that has a similar set of stellar parameters to what may be expected for UDGs (i.e. old, metal poor). We then smooth these templates to the instrumental resolution of KCWI with added smoothing such that the intrinsic dispersion of the template becomes: 0-20 $\mathrm{km\ s^{-1}}$ in steps of 1 $\mathrm{km\ s^{-1}}$ and then 20-100 $\mathrm{km\ s^{-1}}$ in steps of 10 $\mathrm{km\ s^{-1}}$. After smoothing we add Gaussian noise such that the S/N of the resulting template becomes 10/15/20.

We show one example of the above test in Figure \ref{fig:SNtest}. Here we test with a template selected from the ELODIE library, smoothed to an intrinsic velocity dispersion of 0-20 $\mathrm{km\ s^{-1}}$ in steps of 1 $\mathrm{km\ s^{-1}}$ and with added noise until the resulting S/N was 15. We then fit the resulting mock spectra with the \citet{Coelho2014} and ELODIE libraries and display the difference between what \texttt{pPXF} extracts ($\sigma_{pPXF}$) and the intrinsic velocity dispersion inserted into the ELODIE template ($\sigma_{true}$).

Immediately apparent in Figure \ref{fig:SNtest} is the general decrease in precision, from fitting the mock data with either stellar library, when moving towards low intrinsic velocity dispersions. We note that there seems to be a slight coincidence between this decrease in precision and the resolution of the templates being used to fit. Namely, when fitting with the \citet{Coelho2014} library ($\sigma_{coelho} \sim 6$ $\mathrm{km\ s^{-1}}$) errors in the measured velocity dispersion increase below this value. Similarly, the ELODIE library ($\sigma_{ELODIE} \sim 12$ $\mathrm{km\ s^{-1}}$) has great difficulty precisely recovering a velocity dispersion below its own resolution. Similar effects have been reported for the accuracy of MILES library below its resolution \citep{Boardman2016, Boardman2017}.

We note that this source of uncertainty, when trying to recover a velocity dispersion that is poorly sampled by the templates being used to fit, is separate from the systematic error described in section 4 of \citet{Cappellari2017}. Indeed, it is known that large systematic errors can be introduced when trying to recover velocity dispersions when at or below the rate at which data is sampled, which frequently corresponds to the resolution of the data \citep{Robertson2017}. \citet{Cappellari2017} addressed the issue of the velocity dispersion being poorly sampled by the data itself and the ability to overcome this issue using an analytic Fourier transform. While having higher S/N data can assist with recovering velocity dispersions below the templates being used to fit, we note a systematic offset remains even in the near infinite S/N data used in \citet{Cappellari2017}. We suggest that for the lower S/N data likely present for UDGs it is best to use templates of resolution higher than the velocity dispersions that is being recovered.

Beyond the introduction of large errors in smaller velocity dispersions at lower S/N it is difficult to precisely probe the effects of S/N using the three S/N bins of our testing. In general we note that at all intrinsic velocity dispersions the precision is slightly poorer at lower S/N as may be intuitively expected. The effect of S/N on the accuracy of velocity dispersion recovery at sufficiently high intrinsic velocity dispersion (i.e. above the template resolution) is more difficult to draw conclusions from using our testing. Tests using a template from the ELODIE library for the intrinsic spectrum are suggestive of mildly increasing accuracy at higher S/N, while the trend is not as readily apparent in tests using a template from the \citet{Coelho2014}. Here the accuracy of both S/N $=$ 15 and S/N $=$ 20 testing appears similar. We caution however that, having only a singular realisation of each low S/N spectrum, we are vulnerable to statistical effects in our testing of low S/N when comparing our three S/N bins in this manner. In all cases the accuracy of recovery at a velocity dispersion similar to VCC 1287 is within uncertainties quoted in the main body of this work.

Additionally apparent in Figure \ref{fig:SNtest} are the systematically lower velocity dispersions reported by \texttt{pPXF} when fitting using the ELODIE library. In choosing the template to create the mock test we had only two options for stars of similar low metallicity ([Fe/H] $\sim$ $-$1.5) and spectral type that may be characteristic of a UDG. In selecting one of these to build the mock data, and excluding it for the resulting fitting, we are only left with one template in a similar region of stellar parameter space, which may not accurately reflect our mock data. We suggest it is likely that it, along with the other templates in the ELODIE library, do not accurately reflect the mock data and that these systematically lower velocity dispersions are the result of template mismatch in our fitting.

\subsection{\texttt{pPXF} Input Parameters}

In order to test the effects of differing input parameters, namely the number of Gauss--Hermite moments fitted and the additive/multiplicative polynomial used in conjunction with the fitting template, we run \texttt{pPXF} over a wide range of these parameters. We fit our spectra with either 0 or 2 Gauss--Hermite moments (i.e. pure Gaussians or those with $\mathrm{h_3}$ and $\mathrm{h_4}$ also) and between 0 to 10 additive/multiplicative Legendre polynomials resulting in 242 differing \texttt{pPXF} parameter combinations. Additional to this we perform the fitting of our spectra with the synthetic \citet{Coelho2014} stellar library (R $\sim$ 20000), the empirical stellar library ELODIE \citep{ELODIE2004} (R $\sim$ 10000) and our KCWI observations of the Milky Way GC M3 (R $\sim$ 5000). Of these 242 fits > 98\% of those with the \citet{Coelho2014} library and > 71\% of those with M3 report errors sufficiently small to suggest that the spectrum has been effectively modelled by the templates (i.e. those with errors < 25 $\mathrm{km\ s^{-1}}$ in recessional velocity and velocity dispersion). For both libraries these fits largely converge to a similar answer, within $\sim$3 $\mathrm{km\ s^{-1}}$ in recessional velocity/velocity dispersion. We display the results of fitting using the \citet{Coelho2014} and M3 templates in Figures \ref{fig:corner} and \ref{fig:cornerm3} respectively. We take the median values from the subsequent parameter distributions and use them in Section \ref{sec:RV+SK}.

Conversely, of the 242 fits performed with the ELODIE library < 4 \% report similarly small errors and do not display the same convergence on a singular solution. We are unable to categorically provide an explanation as to why ELODIE is unable to effectively model our data. Based on our previous testing with mock data we speculate that, being an empirical library, the stars in this library do not effectively sample the stellar parameters (e.g. metallicity, alpha enhancement) needed for VCC 1287. Indeed, only $\sim$ 2 stars in the ELODIE library have ages and metallicicites broadly similar to VCC 1287 \citep{Pandya2018} but may not have a similar level of alpha enhancement. Incomplete Hertzsprung-Russell Diagram coverage is a problem known to bias both kinematic and stellar population fitting \citep{Barth2002, Coelho2020}.


\bsp	
\label{lastpage}
\end{document}